\documentclass[preprint,review]{elsarticle}
\usepackage{natbib}
\usepackage{graphicx} 
\usepackage{amsmath,amssymb}
\usepackage[margin=1.0in]{geometry}
\usepackage{tikz}
\usetikzlibrary{positioning}
\usepackage{arydshln}
\usepackage[page]{appendix}

\begin{document}

\title{Asymptotic constraints for 1D planar grey photon diffusion from linear transport
with special-relativistic effects}
\author[1]{Ryan T.~Wollaeger}
\author[2]{Jim E.~Morel}
\author[1]{Kendra P.~Long}
\author[1]{Mathew A.~Cleveland}
\author[1]{Robert B.~Lowrie}

\affiliation[1]{organization={Computational Physics and Methods, Los Alamos National Laboratory},
            addressline={P.O. Box 1663}, 
            city={Los Alamos},
            postcode={87545}, 
            state={NM},
            country={USA}}
\affiliation[2]{organization={Department of Nuclear Engineering, Texas A\&M University},
            addressline={423 Spence Street},
            city={College Station},
            postcode={77843},
            state={TX},
            country={USA}}

\begin{abstract}
We derive a grey linear diffusion equation for photons with respect to inertial (or lab-frame)
space and time, using asymptotic analysis in 1D planar geometry.
The solution of the equation is the comoving radiation energy density.
Our analysis does not make use of assumptions about the magnitude of velocity;
instead we derive an asymptotic scaling in the lab frame such that we avoid apparent non-physical
pathologies that are encountered with the standard static-matter scaling.
We permit the photon direction to be continuous (as opposed to constraining the analysis to
discrete ordinates).
The result is a drift-diffusion equation in the lab frame for comoving radiation energy density,
with an adiabatic term that matches the standard semi-relativistic diffusion equation.
Following a recent study for discrete directions, this equation reduces to a pure advection
equation as the velocity approaches the speed of light.
We perform preliminary numerical experiments comparing solutions to relativistic lab-frame
Monte Carlo transport and to the well-known semi-relativistic diffusion equation.
\end{abstract}

\maketitle

\section{Introduction}

Radiation (photon, neutrino) diffusion through optically thick matter moving at
(near-)relativistic speed or in the presence of gravity effects is a ubiquitous phenomenon in
the study of astrophysical plasmas.
It is well known to be a limiting behavior of the more general phenomenon of radiative transfer.
For instance, photon propagation through expanding supernova ejecta can be characterized
with a diffusion solution (see, for instance,~\cite{arnett1982,pinto2000}).
Numerical implementation of the diffusion equation can be used to accelerate
radiative transfer in optically thick regions of space and frequency.
One example is Monte Carlo (MC) accelerated with random walk (RW) or discrete diffusion Monte
Carlo (DDMC)
\cite{fleck1984,gentile2001,densmore2007,densmore2012,cleveland2014,keady2017,smith2020}.
These RW and DDMC methods have been developed for applications to: neutrinos in core collapse
supernova engines~\cite{abdikamalov2012}, cosmic rays~\cite{harding2016,fitzaxen2021}, radiation
pressure feedback in stellar cluster dynamics~\cite{tsang2018}, supernova and kilonova spectral
synthesis~\cite{wollaeger2013,wollaeger2014,wagle2023}, and Lyman $\alpha$ photon transfer during
galaxy evolution~\cite{smith2018}.

While diffusion theory has found practical use in describing photon and neutrino fields in
astrophysical plasmas, the infinite signal speed of the diffusion equation is incompatible
with special relativity (see, for instance,~\cite{dunkel2007,gzyl2025}).
This can readily be seen for static material backgrounds, where the Green's function conforms
to a Gaussian distribution, which has non-zero values everywhere.
As a result, the applicability of diffusion is limited to high-optical depth regions, unless
the equation is modified to restrict the signal speed (for instance $P_1$, $M_1$~\cite{levermore1984} and flux-limit diffusion~\cite{levermore1981}).

It is straightforward to show that the diffusion equation is not Lorentz-invariant.
For instance, consider the 1D planar diffusion equation,
\begin{equation*}
    \frac{\partial\phi_0}{\partial t_0} - D\frac{\partial^2\phi_0}{\partial x_0^2}
    = 0 \;\;,
\end{equation*}
where $t_0$ is time, $x_0$ is the spatial coordinate, $D$ is a constant diffusion
coefficient, and $\phi_0$ is the particle density.
Applying the standard Lorentz transformations for the partial derivatives
(see, for instance,~\cite{lowrie2023}) gives
\begin{equation*}
    \gamma\left(\frac{\partial}{\partial t}
    + v\frac{\partial}{\partial x}\right)\phi_0 
    - D\left[
    \gamma\left(\frac{\partial}{\partial x}+\frac{v}{c^2}\frac{\partial}{\partial t}\right)
    \right]^2\phi_0
    = 0 \;\;,
\end{equation*}
where $v$ is some velocity along $x$, $c$ is the speed of light in vacuum, 
$\gamma=1/\sqrt{1-\beta^2}$ and $\beta=v/c$.
Assuming $v$ is constant in $x$ and $t$, this equation can be expanded to
\begin{equation*}
    \left(\frac{\partial}{\partial t}
    + c\beta\frac{\partial}{\partial x}\right)\phi_0 
    - \gamma D\left(\frac{\partial^2}{\partial x^2}
    + 2\frac{\beta}{c}\frac{\partial^2}{\partial t\partial x}
    + \frac{\beta^2}{c^2}\frac{\partial^2}{\partial t^2}\right)\phi_0
    = 0 \;\;.
\end{equation*}
This resembles a standard drift-diffusion equation if the second derivatives
involving $t$ are neglected.

The problem of unifying diffusion with the principles of special relativity has been
recognized in some form for approximately one century; some approaches to combining them
include: (i) statistical analysis of stochastic kinematics and (ii) asymptotic
analysis of the radiative transfer equation with relativistic corrections.
The latter approach is often pursued in transport theory and astrophysics literature,
where the effects of collective fluid motion on particle distributions is a focus.

Stochastic analyses originated by~\cite{hakim1965} and~\cite{dudley1966} (and further
developed in subsequent works) demonstrate that relativity breaks
Markovian processes in space, where one would expect that the set of spatial
positions for the current particle state completely determine the next state
\cite{dudley1966,dunkel2007,giona2017}.
In developing the relativistic theory, one might first consider bounding the particle
speed in a Poisson-Kac process~\cite{kac1974}, which corresponds to the Telegrapher ($P_1$)
equation and is non-Markovian~\cite{kac1974,dunkel2007}.
Spatially non-Markovian stochastic processes have been formulated in special
\cite{dunkel2007,serva2020} and general~\cite{herrmann2010} relativistic settings.
Properly accounting for relativity in a collective sense, where a frame
is moving relative to the frame in which diffusion is a good description, is a somewhat
distinct problem.
To this end, \cite{giona2017} examine the particle motion as a Poisson-Kac process
along discrete directions, where the process exists in its usual form in one inertial
(comoving) frame, finding in 1D that the effective diffusion coefficient in another
inertial frame is $1/\gamma^3$ times the comoving coefficient; \cite{giona2017} observe
that this follows the intuition that 
$D\sim \delta x^2/\delta t \sim (\delta x_0/\gamma)^2/(\gamma\delta t_0)$,
from Lorentz contraction of space and time dilation.

For asymptotic analysis, the principal assumption is typically that the inverse of the
macroscopic opacity is large compared to all length scales of interest~\cite{thomas1930}.
Unlike the works above that focus on the statistical mechanics of the particles, this
approach usually: starts from the full radiative transfer equation, imposes
the assumed parameter scalings, and expands the radiation intensity in powers of a small-scale
parameter (sometimes labeled $\varepsilon$).
Asymptotic analysis has been applied to deriving a semi-relativistic diffusion equation
\cite{morel2006,lowrie2014}, where $\beta$ is assumed to be asymptotically small (equivalently: 
the hydrodynamic flow time scale is assumed much longer than the light crossing time at the
length scale of interest).
In the fully relativistic context, including general relativity, asymptotic analysis has been
examined early on by~\cite{thomas1930} and subsequently generalized by~\cite{anderson1972}
(see also~\cite{achterberg2018}).
Building off of these works, \cite{shibata2011} demonstrate the truncated Thorne moment
formalism (see~\cite{thorne1981}) converges to a diffusion equation
with modifications from fluid acceleration and shear in the optically thick limit.
The analyses of~\cite{anderson1972} and~\cite{shibata2011} consider both asymptotically large
absorption and scattering opacity.
This is consistent with the static-material scaling that furnishes equilibrium diffusion from
non-linear thermal radiative transfer,
but departs in not assuming that the light crossing distance on the time scale of interest is
large compared to the spatial domain of interest~\cite{larsen1983}.

In the present study, we instead consider an asymptotic analysis that restricts to a special
relativistic, 1D-planar, linear transport context, and we focus on the correspondence of the
lab frame (some inertial frame) to the non-inertial frame comoving with fluid parcels of the
background matter.
In a departure from the work of~\cite{anderson1972} and~\cite{shibata2011} we attempt to use
the standard parameter scaling for linear transport (see, for instance,
\cite{habetler1975,larsen1992}) in the comoving frame.
Thus the scattering opacity is assumed to be asymptotically large while the absorption opacity
is assumed to be asymptotically small, and there is not an equilibrium solution (for instance,
the Planck or Fermi-Dirac distributions in the comoving frame).
In doing so, our objective is to: (i) highlight apparent non-physical pathologies that
occur from applying this standard scaling when velocity gradients are present,
(ii) demonstrate that a particular scaling furnishes the Poisson-Kac-derived $\gamma$-dressed 
result of~\cite{giona2017}, but for continuous directions and including the standard
grey contribution from adiabatic expansion and Doppler shift (see, for instance
\cite{castor2004}), (iii) compare solutions of the diffusion equation to a
relativistic linear transport solution and better-known semi-relativistic diffusion equation
\cite{castor2004}.

This paper is organized as follows.
In Section~\ref{sec:deriv}, we present all considerations for deriving the fully relativistic
1D planar grey diffusion equation, starting from a linear transport equation with isotropic
opacity and elastic isotropic scattering in the comoving frame.
This includes considering the effect of frequency Doppler shift on integration over frequency
for the lab-frame grey equation, in Section~\ref{sec:grey}.
We present properties of a simple class of functions, which are weighted powers of Lorentz
transform (Doppler shift) factors of frequency, and their associated integrals in
Section~\ref{sec:lamb}.
These functions and their properties facilitate both the asymptotic derivation and
the lab-frame harmonic expansion (Appendix~\ref{sec:app1},~\ref{sec:app2}).
In Section~\ref{sec:asym}, we present the development of the lab-frame asymptotic scaling
that furnishes a closed comoving equation with respect to lab-frame space-time coordinates.
This section is subdivided into sections that showcase the pathologies encountered from simpler
asymptotic scaling attempts.
In Section~\ref{sec:disc}, we present a discretization of the fully relativistic diffusion
equation, identifying terms with corresponding non- or semi-relativistic DDMC analogs.
We then compare this discretization to lab-frame MC transport to provide a preliminary
exploration of where the fully relativistic diffusion equation is applicable in Section
\ref{sec:numres}.
In the appendix, we provide: (A) details of the integrated Doppler function recursion, (B) the
aforementioned harmonic ($P_1$) expansion, and finally (C) a brief non-relativistic attempt
of the apparent opacity-only scaling of~\cite{thomas1930,anderson1972,shibata2011}.

\section{Derivation of 1D planar fully relativistic grey diffusion}
\label{sec:deriv}

\subsection{Grey transport with isotropic comoving scattering}
\label{sec:grey}

Assuming 1D planar, grey, linear transport in the inertial lab frame,
\begin{equation}
    \label{eq1:grey}
    \frac{1}{c}\frac{\partial\psi}{\partial t} + \mu\frac{\partial \psi}{\partial x}
    + \sigma_t\psi = \int_{-1}^{1}R_s(\mu'\rightarrow\mu)\psi'd\mu' + \frac{q}{2} \;\;,
\end{equation}
where $c$ is the speed of light, $t$ is time, $x$ is position, $\mu$ is the x-component of
direction, $\sigma_t$ is the lab-frame total opacity, $R_s(\mu'\rightarrow\mu)$ is the
scattering redistribution kernel for angle, $q$ is a lab-frame source term, and $\psi$ is
the lab-frame intensity.
In subsequent sections, we will also use
\begin{subequations}
    \begin{gather}
        \phi = cE = \int_{-1}^1\psi \,d\mu \;\;,\\
        F = \int_{-1}^1\mu\psi\,d\mu \;\;,\\
        P = \int_{-1}^1\mu^2\psi\,d\mu \;\;,
    \end{gather}
\end{subequations}
where $\psi$, $E$, $F$ and $P$ are the scalar intensity, energy density, flux, and pressure
of the radiation field.
Comoving frame versions of all defined quantities will be subscripted with $0$.

We first derive the lab-frame scattering kernel under the assumption of frequency-independent,
isotropic, elastic scattering in the comoving frame.
Frequency and the full direction vector, $\hat{\Omega}=(\mu, \eta, \xi)$, are included to more
readily use Lorentz transform invariants.
Neglecting spatial gradients and sources, the comoving frame equation is
\begin{equation}
    \label{eq2:grey}
    \sigma_{t,0}\psi_0(\nu_0,\hat{\Omega}_0) 
    = \sigma_{s,0}\int_{4\pi}\int_0^{\infty}
    \frac{\delta(\nu_0'-\nu_0)}{4\pi}\psi_0(\nu_0',\hat{\Omega}_0')d\nu_0' d\Omega_0' \;\;,
\end{equation}
where $\delta(\cdot)$ is the Dirac delta distribution.
If the opacity is purely scattering, $\sigma_{t,0}=\sigma_{s,0}$, integrating Eq.~\eqref{eq2:grey}
over solid angle would give $0=0$, indicating conservation.
Making use of the Lorentz invariants and transformation,
\begin{equation}
    \label{eq3:grey}
    \left(\frac{\nu_0}{\nu}\right)^2\sigma_t\psi(\nu,\hat{\Omega}) 
    = \frac{1}{4\pi}\sigma_{s,0}\int_{4\pi}\int_0^{\infty}
    \delta\left(\nu'\gamma(1-\hat{\Omega}'\cdot\vec{\beta})
    - \nu\gamma(1-\hat{\Omega}\cdot\vec{\beta})\right)
    \left(\frac{\nu_0'}{\nu'}\right)^3\psi(\nu',\hat{\Omega}')
    \left(\frac{\nu'}{\nu_0'}\right)d\nu' d\Omega' \;\;,
\end{equation}
where $\vec{\beta}$ is velocity divided by the speed of light, and $\gamma=1/\sqrt{1-\beta^2}$.
The ratios $\nu_0/\nu=\gamma(1-\vec{\beta}\cdot\hat{\Omega})$ are frequency-independent.
Simplifying,
\begin{multline}
    \label{eq4:grey}
    \left(\gamma(1-\vec{\beta}\cdot\hat{\Omega})\right)^2\sigma_t\psi(\nu,\hat{\Omega}) = 
    \\
    \frac{1}{4\pi}\sigma_{s,0}
    \int_{4\pi}\left(\gamma(1-\vec{\beta}\cdot\hat{\Omega}')\right)^2
    \left\{\int_0^{\infty}
    \delta\left(\nu'\gamma(1-\hat{\Omega}'\cdot\vec{\beta})
    - \nu\gamma(1-\hat{\Omega}\cdot\vec{\beta})\right)
    \psi(\nu',\hat{\Omega}')
    d\nu'\right\} d\Omega' \;\;.
\end{multline}
Using the following substitution pattern,
\begin{equation*}
    \int_0^{\infty}\delta(A\nu'-B\nu)f(\nu')d\nu'
    = \frac{1}{A}\int_0^{\infty}\delta(u'-B\nu)f(u'/A)du'
    = \frac{1}{A}f\left(\frac{B}{A}\nu\right)
\end{equation*}
the inner integral over pre-scatter lab-frame frequency can be simplified,
\begin{multline}
    \label{eq5:grey}
    \left(\gamma(1-\vec{\beta}\cdot\hat{\Omega})\right)^2\sigma_t\psi(\nu,\hat{\Omega}) = 
    \frac{1}{4\pi}\sigma_{s,0}
    \int_{4\pi}\left(\gamma(1-\vec{\beta}\cdot\hat{\Omega}')\right)^2
    \left\{\frac{1}{\gamma(1-\vec{\beta}\cdot\hat{\Omega}')}
    \psi\left(\left(\frac{1-\vec{\beta}\cdot\hat{\Omega}}{1-\vec{\beta}\cdot\hat{\Omega}'}\right)\nu, \hat{\Omega}'\right)
    \right\} d\Omega'
    \\
    = \frac{\gamma}{4\pi}\sigma_{s,0}
    \int_{4\pi}(1-\vec{\beta}\cdot\hat{\Omega}')
    \psi\left(\left(\frac{1-\vec{\beta}\cdot\hat{\Omega}}{1-\vec{\beta}\cdot\hat{\Omega}'}\right)\nu, \hat{\Omega}'\right) 
    d\Omega'
    \;\;.
\end{multline}
Integrating Eq.~\eqref{eq5:grey} over lab-frame frequency,
\begin{equation}
    \label{eq6}
    \left(\gamma(1-\vec{\beta}\cdot\hat{\Omega})\right)^2\sigma_t\psi(\hat{\Omega}) = 
    \frac{\gamma}{4\pi}\sigma_{s,0}
    \int_{4\pi}(1-\vec{\beta}\cdot\hat{\Omega}')
    \left(\frac{1-\vec{\beta}\cdot\hat{\Omega}'}{1-\vec{\beta}\cdot\hat{\Omega}}\right)
    \psi(\hat{\Omega}')d\Omega'
    \;\;,
\end{equation}
where we have dropped the $\nu$ argument in $\psi$,
\begin{equation*}
    \psi(\hat{\Omega}) = \int_0^{\infty}\psi(\nu,\hat{\Omega})d\nu \;\;.
\end{equation*}
The factor multiplying $\psi(\hat{\Omega}')$ comes from the following substitution,
\begin{equation*}
    \int_0^{\infty}f\left(\frac{B}{A}\nu\right)d\nu
    = \frac{A}{B}\int_0^{\infty}f(u')du' \;\;.
\end{equation*}
Further simplifying Eq.~\eqref{eq6},
\begin{equation}
    \label{eq7:grey}
    \sigma_t\psi(\hat{\Omega}) = 
    \frac{1}{4\pi}
    \left(\frac{\sigma_{s,0}}{\gamma(1-\vec{\beta}\cdot\hat{\Omega})^3}\right)
    \int_{4\pi}(1-\vec{\beta}\cdot\hat{\Omega}')^2
    \psi(\hat{\Omega}')d\Omega'
    \;\;.
\end{equation}
Assuming velocity $\vec{\beta}=\beta\hat{x}$, where $\hat{x}$ is the unit vector in the
$x$-direction, and $\psi$ is symmetric under rotation about the $x$-axis, integrating over
the azimuthal factor of solid angle gives
\begin{equation}
    \label{eq8:grey}
    \sigma_t\psi(\mu) = 
    \frac{1}{2}
    \left(\frac{\sigma_{s,0}}{\gamma(1-\beta\mu)^3}\right)
    \int_{-1}^1(1-\beta\mu')^2\psi(\mu')d\mu'
    \;\;.
\end{equation}
Equation~\eqref{eq8:grey} implies
\begin{equation*}
    R_s(\mu'\rightarrow\mu) = \frac{1}{2}
    \left(\frac{\sigma_{s,0}}{\gamma(1-\beta\mu)^3}\right)(1-\beta\mu')^2
\end{equation*}
in Eq.~\eqref{eq1:grey}, which would become
\begin{equation}
    \label{eq9:grey}
    \frac{1}{c}\frac{\partial\psi}{\partial t} +
    \mu\frac{\partial \psi}{\partial x} + \sigma_t\psi = 
    \frac{1}{2}\left(\frac{\sigma_{s,0}}{\gamma(1-\beta\mu)^3}\right)
    \int_{-1}^1(1-\beta\mu')^2\psi(\mu')d\mu' + \frac{q}{2} \;\;,
\end{equation}
or
\begin{equation}
    \label{eq10:grey}
    \frac{1}{c}\frac{\partial\psi}{\partial t} +
    \mu\frac{\partial \psi}{\partial x} + \sigma_{t,0}\gamma(1-\beta\mu)\psi = 
    \frac{1}{2}\left(\frac{\sigma_{s,0}}{\gamma(1-\beta\mu)^3}\right)
    \int_{-1}^1(1-\beta\mu')^2\psi(\mu')d\mu' + \frac{q}{2} \;\;.
\end{equation}
For $\beta=0$, Eq.~\eqref{eq10:grey} reduces to the usual grey transport equation with isotropic
scattering.

\subsection{$\lambda$ function and $\Lambda$ integral}
\label{sec:lamb}

For convenience, we introduce the class of functions,
\begin{equation}
    \label{eq1:lamb}
    \lambda_{n,k}(\mu) = \frac{\mu^k}{\gamma^n(1-\beta\mu)^n} \;\;.
\end{equation}
We find that using and reasoning about these functions in the context of the following
sections and appendix expedites evaluation of relativistic factors.
Here we review the properties of this class of functions through multiplication,
differentiation in space or time, and integration over $\mu$; these operations are relevant
to the following sections and appendix.

We note Eq.~\eqref{eq1:lamb} readily satisfies a simple product rule that can be useful for
keeping track of multiple Lorentz transforms in products of terms in harmonic ($\mu^k$-weighted)
expansions,
\begin{equation}
    \label{eq2:lamb}
    \lambda_{n+m,k+l}(\mu) = \lambda_{n,k}(\mu)\lambda_{m,l}(\mu) \;\;.
\end{equation}
Differentiation in space or time can be posed as a recursion,
\begin{equation}
    \label{eq3:lamb}
    \frac{\partial\lambda_{n,k}}{\partial x}
    = \mu^k\left(\gamma^{-n}\frac{\partial(1-\beta\mu)^{-n}}{\partial x}
    + (1-\beta\mu)^{-n}\frac{\partial\gamma^{-n}}{\partial x}\right)
    = n\gamma\frac{\partial\beta}{\partial x}
    \left(\lambda_{n+1,k+1}(\mu) - \gamma\lambda_{n,k}(\mu)\right) \;\;,
\end{equation}
where we have made use of
\begin{equation}
    \label{eq4:lamb}
    \frac{\partial\gamma}{\partial x} = \gamma^3\frac{\partial\beta}{\partial x} \;\;.
\end{equation}
In Section~\ref{sec:asym} we observe occurrences of
\begin{equation}
    \label{eq5:lamb}
    \frac{\partial\ln\lambda_{n,k}}{\partial x} 
    = \frac{1}{\lambda_{n,k}}\frac{\partial\lambda_{n,k}}{\partial x}
    = n\gamma\frac{\partial\beta}{\partial x}
    \left(\lambda_{1,1}(\mu) - \gamma\right) \;\;,
\end{equation}
which can be seen by applying Eq.~\eqref{eq2:lamb}.
We caution that derivatives of $\lambda_{n,k}$ with respect to space or time
are more complicated to evaluate in terms of $\mu_0$, given $\mu_0$ varies in space
and time (see Section~\ref{sec:abbr}).

We further introduce integrals of $\lambda_{n,k}(\mu)$ over $\mu$,
\begin{equation}
    \Lambda_{n,k} = \int_{-1}^1\lambda_{n,k}(\mu)d\mu \;\;.
\end{equation}
It is straightforward to show (Appendix Section~\ref{sec:app1}) that the $\Lambda_{n,k}$
values follow a triangular recursion relationship,
\begin{equation}
    \label{eq7:lamb}
    \Lambda_{n,k} = \frac{1}{\beta}\Lambda_{n,k-1} - \frac{1}{\gamma\beta}\Lambda_{n-1,k-1}
    \;\;.
\end{equation}
We may use this recursion to efficiently evaluate the 1D planar lab-frame
harmonic P$_n$-series expansion of Eq.~\eqref{eq10:grey}, as shown in the Appendix.

By virtue of the Lorentz transform for frequency, we also note the relation,
\begin{equation}
    \label{eq8:lamb}
    \lambda_{n,0}(\mu) = \frac{1}{\gamma^n(1-\beta\mu)^n} 
    = \gamma^n(1+\beta\mu_0)^n = \frac{1}{\lambda_{n,0}(-\mu_0)} \;\;,
\end{equation}
and for $k>0$,
\begin{equation}
    \label{eq9:lamb}
    \lambda_{n,k}(\mu) = \frac{\mu^k}{\gamma^n(1-\beta\mu)^n} 
    = \frac{(\mu_0+\beta)^k}{(1+\beta\mu_0)^k}\gamma^n(1+\beta\mu_0)^n 
    = \gamma^k\sum_{q=0}^k\binom{k}{q}\frac{(-1)^q\beta^{k-q}}{\lambda_{n-k,-q}(-\mu_0)} \;\;.
\end{equation}
We use Eqs.~\eqref{eq8:lamb} and~\eqref{eq9:lamb} in Section~\ref{sec:asym} to change
reference frames before integration over $\mu_0$.

\subsection{Angular derivative in the comoving frame}
\label{sec:abbr}

The approaches to asymptotic scaling makes use of the comoving frame.
As such, it is worth noting that the space or time derivative of the comoving intensity,
$\psi_0$, must account for the change in $\mu_0$ with space or time.
Furthermore, from the Lorentz invariant relation of $\psi$ to $\psi_0$,
\begin{equation}
    \label{eq1:abbr}
    \psi(x,\mu) = \frac{\psi_0(x,\mu_0)}{\gamma^4(1-\mu\beta)^4}
    = \lambda_{4,0}(\mu)\psi_0(x,\mu_0) \;\;.
\end{equation}
Using the angular derivative, the relationship between spatial derivatives is
\begin{equation}
    \label{eq2:abbr}
    \frac{\partial\psi}{\partial x} = \frac{\partial}{\partial x}
    \left(\lambda_{4,0}\psi_0\right)
    = \psi_0\frac{\partial\lambda_{4,0}}{\partial x} + \lambda_{4,0}
    \left(\left.\frac{\partial\psi_0}{\partial x}\right|_{\mu_0}
    + \frac{\partial\mu_0}{\partial x}\frac{\partial\psi_0}{\partial\mu_0}\right)
    \;\;,
\end{equation}
where $|_{\mu_0}$ indicates $\mu_0$ is fixed when taking the partial derivative in
$x$ (in order to avoid abuse of notation).
The time derivative can similarly be found, resulting in $x$ replaced with $t$
in Eq.~\eqref{eq2:abbr}.
In 1D planar geometry a simple equation for $\mu_0$ exists in terms of $\mu$ and $\beta$,
\begin{equation}
    \label{eq3:abbr}
    \mu_0 = \frac{\mu - \beta}{1 - \beta\mu} \;\;.
\end{equation}
Consequently, assuming a given space-time dependence of $\beta$,
\begin{equation}
    \label{eq4:abbr}
    \frac{\partial\mu_0}{\partial x} 
    = -\frac{(1-\mu^2)}{(1 - \beta\mu)^2}\frac{\partial\beta}{\partial x} 
    = -\gamma^2(1-\mu_0^2)\frac{\partial\beta}{\partial x} \;\;,
\end{equation}
where we have used the fact that $\mu$ is constant in space and time in the lab
frame, for a particular solution $\psi(x,\mu)$.
The time derivative can again be similarly found, resulting in $x$ replaced with
$t$ in Eq.~\eqref{eq4:abbr}.

\subsection{Asymptotic scaling}
\label{sec:asym}

Here we present two scaling methods: in the comoving frame using the
standard approach for static background matter, and with a modification that includes
an anisotropic scalar multiple of $\psi$.
We first attempt to demonstrate the former does not furnish a closed equation to
leading order, then we attempt to show the latter does give a closed drift-diffusion
equation.
In both attempts, we assume the existence of a ``smallness'' parameter $\varepsilon$
that is applicable to both the comoving and lab frame, and that intensity can be expanded
in powers of $\varepsilon$,
\begin{subequations}
    \label{eq1:asym}
    \begin{gather}
        \psi = \sum_{k=0}^{\infty}\psi^{(k)}\varepsilon^k \;\;,\\
        \psi_0 = \sum_{k=0}^{\infty}\psi_0^{(k)}\varepsilon^k \;\;.
    \end{gather}
\end{subequations}
We furthermore assume that Eq.~\eqref{eq1:abbr} holds for each order,
\begin{equation}
    \label{eq2:asym}
    \psi^{(k)}(x,\mu) = \lambda_{4,0}(\mu)\psi_0^{(k)}(x,\mu_0) \;\;,
\end{equation}
which recovers the full intensity invariance.
We must impose Eq.~\eqref{eq2:asym} for $k\in\{0,1,2\}$ in order to undertake the
transformations from the lab to the comoving frame in the following sections.

Concerning the $\varepsilon$-scaling of transport equation, in an equation
with a lab-frame time derivative, we consider scaling of the form
\begin{multline}
    A[\psi] = \frac{1}{c}\frac{\partial\psi}{\partial t} = 
    \frac{1}{c}\frac{\partial\psi}{\partial t}
    + \beta\frac{\partial\psi}{\partial x}
    + B(\mu)\psi
    - \beta\frac{\partial\psi}{\partial x}
    - B(\mu)\psi \\
    \rightarrow
    \varepsilon\left(
    \frac{1}{c}\frac{\partial\psi}{\partial t}
    + \beta\frac{\partial\psi}{\partial x}
    + B(\mu)\psi\right)
    - \beta\frac{\partial\psi}{\partial x} 
    - B(\mu)\psi = A_{\varepsilon}[\psi] \;\;,
\end{multline}
where $A[\cdot]$ is the original time-derivative operator (or the remainder of
the transport equation), $B(\mu)$ is an as-yet unknown coefficient
and $A_{\varepsilon}[\cdot]$ is a $\varepsilon$-scaled form of this operator.
This scaling assumes that the residual excluded from the $\varepsilon$ coefficient
scales as a leading-order contribution.
The full set of scaling relations we use is
\begin{subequations}
    \label{eq4:asym}
    \begin{gather}
        \left(\frac{1}{c}\frac{\partial}{\partial t} + \beta\frac{\partial}{\partial x}
        + B(\mu)\right)\psi \rightarrow
        \varepsilon\left(\frac{1}{c}\frac{\partial}{\partial t}
        + \beta\frac{\partial}{\partial x}
        + B(\mu)\right)\psi \;\;, \\
        \sigma_{t,0} \rightarrow \frac{\sigma_{t,0}}{\varepsilon} \;\;, \\
        \sigma_{a,0} \rightarrow \varepsilon\sigma_{a,0} \;\;, \\
        q \rightarrow \varepsilon q \;\;.
    \end{gather}
\end{subequations}
In the following sections, we determine the form of $B(\mu)$ such that the equation
for the leading-order scalar intensity is closed at O($\varepsilon^2$).

\subsubsection{Scaling the comoving time derivative}

Assuming $B(\mu)=0$, Eqs.~\eqref{eq4:asym} reduce to the standard asymptotic scaling
for linear transport in the diffusion limit, in the comoving frame.
Applying this to Eq.~\eqref{eq1:grey}, and substituting in $\lambda_{n,0}$-functions,
\begin{equation}
    \label{eq5:asym}
    \varepsilon^2\left(\frac{1}{c}\frac{\partial\psi}{\partial t}
    + \beta\frac{\partial\psi}{\partial x}\right)
    + \varepsilon(\mu-\beta)\frac{\partial \psi}{\partial x} 
    + \frac{\sigma_{t,0}}{\lambda_{1,0}}\psi =
    \frac{1}{2}\lambda_{3,0}
    (\sigma_{t,0} - \varepsilon^2\sigma_{a,0})
    \int_{-1}^1\frac{\psi(\mu')}{\lambda_{2,0}(\mu')}d\mu' + \varepsilon^2\frac{q}{2} \;\;,
\end{equation}
where $\lambda_{n,0}=\lambda_{n,0}(\mu)$ for brevity.
Incorporating Eq.~\eqref{eq1:asym} into Eq.~\eqref{eq5:asym} and matching coefficients
of $\varepsilon^k$, the O(1) equation is
\begin{equation}
    \label{eq6:asym}
    \psi^{(0)} = \frac{1}{2}\lambda_{4,0}
    \int_{-1}^1\frac{\psi^{(0)}(\mu')}{\lambda_{2,0}(\mu')}d\mu' \;\;,
\end{equation}
and the O($\varepsilon$) equation is
\begin{equation}
    \label{eq7:asym}
    \frac{(\mu-\beta)}{(1-\beta\mu)}\frac{1}{\gamma\sigma_{t,0}}
    \frac{\partial \psi^{(0)}}{\partial x} 
    + \psi^{(1)} =
    \frac{1}{2}\lambda_{4,0}
    \int_{-1}^1\frac{\psi^{(1)}(\mu')}{\lambda_{2,0}(\mu')}d\mu' \;\;.
\end{equation}
The O(1) equation is a statement of isotropy in the comoving frame, while the
O($\varepsilon$) reduces to the usual linearly anisotropic form when $\beta=0$.

Converting $\mu$, $\psi^{(0)}$ and $\psi^{(1)}$ to the comoving frame
and dividing by a factor of $\lambda_{4,0}(\mu)$, Eqs.~\eqref{eq6:asym}
and~\eqref{eq7:asym} become
\begin{equation}
    \label{eq9:asym}
    \psi_0^{(0)} = \frac{1}{2}\phi_0^{(0)} \;\;,
\end{equation}
and
\begin{equation}
    \label{eq10:asym}
    \frac{1}{\gamma\sigma_{t,0}}
    \mu_0\left(
    \frac{\psi_0^{(0)}}{\lambda_{4,0}}\frac{\partial\lambda_{4,0}}{\partial x} 
    + \left.\frac{\partial\psi_0^{(0)}}{\partial x} \right|_{\mu_0}
    -\gamma^2\frac{\partial\beta}{\partial x}(1-\mu_0^2)
    \frac{\partial\psi_0^{(0)}}{\partial\mu_0}\right)
    + \psi_0^{(1)} = \frac{1}{2}\phi_0^{(1)} \;\;.
\end{equation}
Given $\psi_0^{(0)}$ is isotropic, the angular derivative vanishes and
$|_{\mu_0}$ can be removed from the derivative, leaving
\begin{equation}
    \label{eq10:asym}
    \frac{1}{\gamma\sigma_{t,0}}
    \mu_0\left(
    \psi_0^{(0)}\frac{\partial\ln\lambda_{4,0}}{\partial x} 
    + \frac{\partial\psi_0^{(0)}}{\partial x}\right)
    + \psi_0^{(1)} = \frac{1}{2}\phi_0^{(1)} \;\;.
\end{equation}

If $\beta$ is constant in space, Eq.~\eqref{eq10:asym} reduces to the usual linear
anisotropic form in the comoving frame (consistent with the lab-frame form of the equation).
If $\beta$ is not constant in space, integrating over $\mu_0$ and simplifying gives
\begin{equation}
    \label{eq11:asym}
    \frac{4}{3}\gamma^2\frac{\partial\beta}{\partial x}\phi_0^{(0)} = 0 \;\;,
\end{equation}
which is a contradiction, unless $\phi_0^{(0)}=0$.

\subsubsection{Scaling the comoving time derivative with the anisotropic term}

The contradiction resulting from scaling the comoving time derivative suggests
that we may find a non-zero form of $B(\mu)$ that eliminates the anisotropic term
proportional to the velocity gradient.
Here we show that a form that accomplishes this is
\begin{equation}
    \label{eq12:asym}
    B(\mu) = (\mu - \beta)\frac{\partial}{\partial x}\ln(\lambda_{4,0}(\mu)) \;\;.
\end{equation}
Applying the scaling with Eq.~\eqref{eq12:asym},
\begin{multline}
    \label{eq13:asym}
    \varepsilon^2\left(\frac{1}{c}\frac{\partial\psi}{\partial t}
    + \beta\frac{\partial\psi}{\partial x}
    + \psi\,(\mu - \beta)\frac{\partial\ln\lambda_{4,0}}{\partial x}
    \right) \\
    + \varepsilon(\mu-\beta)\left(\frac{\partial \psi}{\partial x}
    - \psi\,\frac{\partial\ln\lambda_{4,0}}{\partial x}\right)
    + \frac{\sigma_{t,0}}{\lambda_{1,0}}\psi = \\
    \frac{1}{2}\lambda_{3,0}
    (\sigma_{t,0} - \varepsilon^2\sigma_{a,0})
    \int_{-1}^1\frac{\psi(\mu')}{\lambda_{2,0}(\mu')}d\mu' + \varepsilon^2\frac{q}{2} \;\;.
\end{multline}
Incorporating Eq.~\eqref{eq1:asym} into Eq.~\eqref{eq13:asym} and matching coefficients
of $\varepsilon^k$, the O(1) equation is
\begin{equation}
    \label{eq14:asym}
    \psi^{(0)} = \frac{1}{2}\lambda_{4,0}
    \int_{-1}^1\frac{\psi^{(0)}(\mu')}{\lambda_{2,0}(\mu')}d\mu' \;\;,
\end{equation}
and the O($\varepsilon$) equation is
\begin{equation}
    \label{eq15:asym}
    \frac{(\mu-\beta)}{(1-\beta\mu)}\frac{1}{\gamma\sigma_{t,0}}
    \left(
    \frac{\partial \psi^{(0)}}{\partial x}
    - \psi^{(0)}\,\frac{\partial\ln\lambda_{4,0}}{\partial x}
    \right)
    + \psi^{(1)} =
    \frac{1}{2}\lambda_{4,0}
    \int_{-1}^1\frac{\psi^{(1)}(\mu')}{\lambda_{2,0}(\mu')}d\mu' \;\;.
\end{equation}

Converting $\mu$, $\psi^{(0)}$ and $\psi^{(1)}$ to the comoving frame
and dividing by a factor of $\lambda_{4,0}(\mu)$, Eqs.~\eqref{eq14:asym}
and~\eqref{eq15:asym} become
\begin{equation}
    \label{eq9:asym}
    \psi_0^{(0)} = \frac{1}{2}\phi_0^{(0)} \;\;,
\end{equation}
and
\begin{equation}
    \label{eq16:asym}
    \frac{1}{\gamma\sigma_{t,0}}
    \mu_0\left(
    \frac{\psi_0^{(0)}}{\lambda_{4,0}}\frac{\partial\lambda_{4,0}}{\partial x} 
    +\frac{\partial\psi_0^{(0)}}{\partial x}
    - \psi_0^{(0)}\,\frac{\partial\ln\lambda_{4,0}}{\partial x}\right)
    + \psi_0^{(1)} = \frac{1}{2}\phi_0^{(1)} \;\;,
\end{equation}
where we have made use again of isotropy to eliminate the angular derivative.
Simplifying the O($\varepsilon$) equation gives
\begin{equation}
    \label{eq17:asym}
    \frac{1}{\gamma\sigma_{t,0}}
    \mu_0\frac{\partial\psi_0^{(0)}}{\partial x}
    + \psi_0^{(1)} = \frac{1}{2}\phi_0^{(1)} \;\;,
\end{equation}
which is now the standard linear anisotropic relationship, posed in terms of
comoving quantities.
The contradiction encountered by scaling the comoving time derivative is no
longer manifest in Eq.~\eqref{eq17:asym}.

Proceeding to the O($\varepsilon^2$) equation, matching $\varepsilon^2$ coefficients
gives
\begin{multline}
    \label{eq18:asym}
    \lambda_{1,0}\left(\frac{1}{c}\frac{\partial}{\partial t}
    + \beta\frac{\partial}{\partial x}\right)\psi^{(0)}
    + \psi^{(0)}\frac{(\mu-\beta)}{\gamma(1-\beta\mu)}
    \frac{\partial\ln\lambda_{4,0}}{\partial x}
    \\
    + \frac{(\mu-\beta)}{\gamma(1-\beta\mu)}\left(\frac{\partial \psi^{(1)}}{\partial x}
    - \psi^{(1)}\,\frac{\partial\ln\lambda_{4,0}}{\partial x}\right)
    + \sigma_{t,0}\psi^{(2)} = \\
    \frac{1}{2}\lambda_{4,0}\left(
    \sigma_{t,0}\int_{-1}^1\frac{\psi^{(2)}(\mu')}{\lambda_{2,0}(\mu')}d\mu'
    - \sigma_{a,0}\int_{-1}^1\frac{\psi^{(0)}(\mu')}{\lambda_{2,0}(\mu')}d\mu'\right)
    + \lambda_{1,0}\frac{q}{2} \;\;.
\end{multline}
Converting this to the comoving frame and dividing by $\lambda_{4,0}$,
\begin{multline}
    \label{eq19:asym}
    \lambda_{1,0}\left(\frac{1}{c}\frac{\partial}{\partial t}
    + \beta\frac{\partial}{\partial x}\right)\psi_0^{(0)}
    + \lambda_{1,0}\psi_0^{(0)}\left(\frac{1}{c}\frac{\partial}{\partial t}
    + \beta\frac{\partial}{\partial x}\right)\ln\lambda_{4,0}
    + \psi_0^{(0)}\frac{\mu_0}{\gamma}
    \frac{\partial\ln\lambda_{4,0}}{\partial x}
    \\
    + \frac{\mu_0}{\gamma}\left(
    \left.\frac{\partial\psi_0^{(1)}}{\partial x}\right|_{\mu_0} 
    -\gamma^2(1-\mu_0^2)\frac{\partial\beta}{\partial x}
    \frac{\partial \psi_0^{(1)}}{\partial\mu_0}
    + \frac{\psi_0^{(1)}}{\lambda_{4,0}}\frac{\partial\lambda_{4,0}}{\partial x}
    - \psi_0^{(1)}\,\frac{\partial\ln\lambda_{4,0}}{\partial x}\right)
    + \sigma_{t,0}\psi_0^{(2)} = \\
    \frac{1}{2}\left(
    \sigma_{t,0}\phi_0^{(2)}
    - \sigma_{a,0}\phi_0^{(0)}\right)
    + \frac{q}{2\lambda_{3,0}} \;\;,
\end{multline}
where use has been made of $\psi_0^{(0)}$ isotropy in evaluating the first term
on the left side, to eliminate derivaties with respect to $\mu_0$.
We may simplify the $k=1$ terms by canceling the $B(\mu)$ residue and using
Eq.~\eqref{eq17:asym},
\begin{multline}
    \label{eq20:asym}
    \lambda_{1,0}\left(\frac{1}{c}\frac{\partial}{\partial t}
    + \beta\frac{\partial}{\partial x}\right)\psi_0^{(0)}
    + \lambda_{1,0}\psi_0^{(0)}\left(\frac{1}{c}\frac{\partial}{\partial t}
    + \beta\frac{\partial}{\partial x}\right)\ln\lambda_{4,0}
    + \psi_0^{(0)}\frac{\mu_0}{\gamma}
    \frac{\partial\ln\lambda_{4,0}}{\partial x}
    \\
    + \frac{\mu_0}{\gamma}
    \frac{\partial}{\partial x}\left(\frac{\phi_0^{(1)}}{2}\right)
    - \frac{\mu_0}{\gamma}\left.\frac{\partial}{\partial x}
    \left(\frac{\mu_0}{\gamma\sigma_{t,0}}\frac{\partial\psi_0^{(0)}}{\partial x}\right)
    \right|_{\mu_0}
    + \mu_0(1-\mu_0^2)\frac{\partial\beta}{\partial x}
    \frac{1}{\sigma_{t,0}}\frac{\partial\psi_0^{(0)}}{\partial x}
    + \sigma_{t,0}\psi_0^{(2)} = \\
    \frac{1}{2}\left(
    \sigma_{t,0}\phi_0^{(2)}
    - \sigma_{a,0}\phi_0^{(0)}\right)
    + \frac{q}{2\lambda_{3,0}} \;\;.
\end{multline}
Applying Eq.~\eqref{eq5:lamb} to the derivatives of $\ln\lambda_{4,0}$,
\begin{multline}
    \label{eq21:asym}
    \lambda_{1,0}\left(\frac{1}{c}\frac{\partial}{\partial t}
    + \beta\frac{\partial}{\partial x}\right)\psi_0^{(0)}
    + 4\left[\gamma\left(\lambda_{2,1} - \gamma\lambda_{1,0}\right)
    \left(\frac{1}{c}\frac{\partial}{\partial t}
    + \beta\frac{\partial}{\partial x}\right)\beta
    + \mu_0\left(\lambda_{1,1} - \gamma\right)\frac{\partial\beta}{\partial x}
    \right]\psi_0^{(0)}
    \\
    + \frac{\mu_0}{\gamma}
    \frac{\partial}{\partial x}\left(\frac{\phi_0^{(1)}}{2}\right)
    - \frac{\mu_0^2}{\gamma}\frac{\partial}{\partial x}
    \left(\frac{1}{\gamma\sigma_{t,0}}\frac{\partial\psi_0^{(0)}}{\partial x}\right)
    + \mu_0(1-\mu_0^2)\frac{\partial\beta}{\partial x}
    \frac{1}{\sigma_{t,0}}\frac{\partial\psi_0^{(0)}}{\partial x}
    + \sigma_{t,0}\psi_0^{(2)} = \\
    \frac{1}{2}\left(
    \sigma_{t,0}\phi_0^{(2)}
    - \sigma_{a,0}\phi_0^{(0)}\right)
    + \frac{q}{2\lambda_{3,0}} \;\;.
\end{multline}
Now no occurrences of $\lambda$-functions and $\mu_0$ are inside space or time
derivatives.
The source term $q$ may be related to the comoving source term by
\begin{equation*}
    q(\mu) = q_0(\mu_0)\lambda_{3,0}(\mu) \;\;,
\end{equation*}
where $\lambda_{3,0}$ arises from lab-frame frequency integration of the Lorentz
invariant relationship for emissivity.
Substituting $q_0$ for $q$ and using Eq.~\eqref{eq9:lamb} we may convert the
$\lambda$-functions to polynomials in $\mu_0$,
\begin{multline}
    \label{eq22:asym}
    \gamma(1+\beta\mu_0)\left(\frac{1}{c}\frac{\partial}{\partial t}
    + \beta\frac{\partial}{\partial x}\right)\psi_0^{(0)}
    + 4\gamma\left(\mu_0+\beta - 1\right)\left[\gamma^2(1+\beta\mu_0)
    \left(\frac{1}{c}\frac{\partial}{\partial t}
    + \beta\frac{\partial}{\partial x}\right)\beta
    + \mu_0\frac{\partial\beta}{\partial x}
    \right]\psi_0^{(0)}
    \\
    + \frac{\mu_0}{\gamma}
    \frac{\partial}{\partial x}\left(\frac{\phi_0^{(1)}}{2}\right)
    - \frac{\mu_0^2}{\gamma}\frac{\partial}{\partial x}
    \left(\frac{1}{\gamma\sigma_{t,0}}\frac{\partial\psi_0^{(0)}}{\partial x}\right)
    + \mu_0(1-\mu_0^2)\frac{\partial\beta}{\partial x}
    \frac{1}{\sigma_{t,0}}\frac{\partial\psi_0^{(0)}}{\partial x}
    + \sigma_{t,0}\psi_0^{(2)} = \\
    \frac{1}{2}\left(
    \sigma_{t,0}\phi_0^{(2)}
    - \sigma_{a,0}\phi_0^{(0)}\right)
    + \frac{q_0}{2} \;\;.
\end{multline}
Integrating Eq.~\eqref{eq22:asym} over $\mu_0$, dividing by a factor of $\gamma$ and
simplifying gives
\begin{multline}
    \label{eq23:asym}
    \left(\frac{1}{c}\frac{\partial}{\partial t}
    + \beta\frac{\partial}{\partial x}\right)\phi_0^{(0)}
    + 4\left[\gamma^2\left(\frac{4}{3}\beta - 1\right)
    \left(\frac{1}{c}\frac{\partial}{\partial t}
    + \beta\frac{\partial}{\partial x}\right)\beta
    + \frac{1}{3}\frac{\partial\beta}{\partial x}
    \right]\phi_0^{(0)}
    \\
    - \frac{1}{3\gamma^2}\frac{\partial}{\partial x}
    \left(\frac{1}{\gamma\sigma_{t,0}}\frac{\partial\phi_0^{(0)}}{\partial x}\right)
     = -\frac{\sigma_{a,0}}{\gamma}\phi_0^{(0)}
    + \frac{1}{2\gamma}\int_{-1}^1q_0(\mu_0)d\mu_0 \;\;.
\end{multline}

We see that we have obtained a drift-diffusion equation for the leading-order scalar
intensity using our lab-frame asymptotic scaling with $B(\mu)$ from Eq.~\eqref{eq12:asym}.
Furthermore, we note that we have not assumed a constraint on the value of $\beta$ itself.
The diffusion stencil has three factors of $1/\gamma$, consistent with the finding of
\cite{giona2017} for bi-directional 1D planar random walk.
However, we have an additional $\beta$-dependent coefficient of $\phi_0^{(0)}$, which
accounts for the energetic effect of Doppler shift and adiabatic expansion.
To more clearly see this, we may neglect the time-dependence of $\beta$, and neglect
terms of O($\beta^2$) and O($\beta\partial\beta/\partial x$), to obtain
\begin{equation}
    \label{eq24:asym}
    \left(\frac{1}{c}\frac{\partial}{\partial t}
    + \beta\frac{\partial}{\partial x}\right)\phi_0^{(0)}
    + \frac{4}{3}\frac{\partial\beta}{\partial x}\phi_0^{(0)}
    - \frac{1}{3}\frac{\partial}{\partial x}
    \left(\frac{1}{\sigma_{t,0}}\frac{\partial\phi_0^{(0)}}{\partial x}\right)
     = -\sigma_{a,0}\phi_0^{(0)} + \frac{1}{2}\int_{-1}^1q_0(\mu_0)d\mu_0 \;\;,
\end{equation}
which is the standard semi-relativistic form (see, for instance,~\cite{castor2004}, Chapter 6).
If $\beta$ is constant and approaches 1, assuming $q$ and $\phi_0^{(0)}$ remain bounded,
\begin{equation}
    \label{eq25:asym}
    \left(\frac{1}{c}\frac{\partial}{\partial t}
    + \beta\frac{\partial}{\partial x}\right)\phi_0^{(0)}
     = 0 \;\;,
\end{equation}
which is the limiting behavior observed by~\cite{giona2017}.

\subsubsection{Scaling the Lagrangian time derivative of $\beta$}

We note that there is pathology in Eq.~\eqref{eq23:asym}: the coefficient accounting for
adiabatic and Doppler shift has a factor that is mixed-parity in $\beta$, $(4\beta/3-1)$,
multiplied to the Lagrangian derivative of $\beta$.
This combination of factors breaks symmetry under the parity transformation
$(x,\beta)\rightarrow(-x,-\beta)$, which is maintained in the semi-relativistic form of
Eq.~\eqref{eq24:asym}.
To preserve this physical symmetry, this suggests we must impose an asymptotic
non-acceleration condition for the matter, for instance,
\begin{equation}
    \label{eq26:asym}
    \left(\frac{1}{c}\frac{\partial}{\partial t}
    + \beta\frac{\partial}{\partial x}\right)\beta
    \rightarrow \varepsilon \left(\frac{1}{c}\frac{\partial}{\partial t}
    + \beta\frac{\partial}{\partial x}\right)\beta \;\;.
\end{equation}
Incorporating Eq.~\eqref{eq26:asym} into the suite of scaling conditions removes
the term corresponding to the Lagrangian derivative of velocity in the O($\varepsilon^2$)
equation, resulting in
\begin{equation}
    \label{eq27:asym}
    \left(\frac{1}{c}\frac{\partial}{\partial t}
    + \beta\frac{\partial}{\partial x}\right)\phi_0^{(0)}
    + \frac{4}{3}\frac{\partial\beta}{\partial x}\phi_0^{(0)}
    - \frac{1}{3\gamma^2}\frac{\partial}{\partial x}
    \left(\frac{1}{\gamma\sigma_{t,0}}\frac{\partial\phi_0^{(0)}}{\partial x}\right)
     = -\frac{\sigma_{a,0}}{\gamma}\phi_0^{(0)} + \frac{Q_0}{\gamma} \;\;,
\end{equation}
where
\begin{equation*}
    Q_0 \equiv \frac{1}{2}\int_{-1}^1q_0(\mu_0)d\mu_0 \;\;.
\end{equation*}
This equation now has the same coefficient for adiabatic and expansion effects as the
standard semi-relativistic form, so retains mirrored behavior in scalar intensity when
$(x,\beta)\rightarrow(-x,-\beta)$.
In subsequent sections we examine Eq.~\eqref{eq27:asym}, which corresponds to the
following list of lab-frame scaling relations, now including Eq.~\eqref{eq26:asym},
\begin{subequations}
    \label{eq28:asym}
    \begin{gather}
        \left(\frac{1}{c}\frac{\partial}{\partial t} + \beta\frac{\partial}{\partial x}
        + (\mu - \beta)\frac{\partial}{\partial x}\ln(\lambda_{4,0}(\mu))\right)\psi \rightarrow
        \varepsilon\left(\frac{1}{c}\frac{\partial}{\partial t}
        + \beta\frac{\partial}{\partial x}
        + (\mu - \beta)\frac{\partial}{\partial x}\ln(\lambda_{4,0}(\mu))\right)\psi \;\;, \\
        \sigma_{t,0} \rightarrow \frac{\sigma_{t,0}}{\varepsilon} \;\;, \\
        \sigma_{a,0} \rightarrow \varepsilon\sigma_{a,0} \;\;, \\
        q \rightarrow \varepsilon q \;\;, \\
        \left(\frac{1}{c}\frac{\partial}{\partial t}
        + \beta\frac{\partial}{\partial x}\right)\beta
        \rightarrow \varepsilon \left(\frac{1}{c}\frac{\partial}{\partial t}
        + \beta\frac{\partial}{\partial x}\right)\beta \;\;.
    \end{gather}
\end{subequations}

\section{Discretization}
\label{sec:disc}

\subsection{Spatial stencil}

Here we describe the discretization of Eq.~\eqref{eq27:asym} we use to explore properties
of the solutions in Section~\ref{sec:numres}.
We solve the resulting discretized system deterministically, but we note that there
exists a straightforward Discrete Diffusion Monte Carlo (DDMC)
\cite{gentile2001,densmore2007,abdikamalov2012} interpretation of the discretized terms, and
write them following the notation of~\cite{densmore2007}.
In 1D, a grey deterministic solution is straightforward to implement, and we do not seek to
hybridize lab-frame MC transport with DDMC in the scope of this work.

We make the assumption that $\beta$ and the spatial derivatives of $\beta$ are
evaluated per spatial cell, prior to spatial discretization of Eq.~\eqref{eq27:asym}.
Similar to opacity, we assume $\beta$ and its derivatives are given with a time step
and spatial cell.
To simplify the presentation of the spatial discretization that follows, we introduce
a label for the coefficient of $\phi_0^{(0)}$ on the left side,
\begin{equation}
    \eta_i \equiv \frac{4}{3}\left.\frac{\partial\beta}{\partial x}\right|_i
\end{equation}
where subscript $i$ is the index of the spatial cell, indicating evaluation at the cell.

Finite-volume discretization of Eq.~\eqref{eq27:asym} gives
\begin{multline}
    \label{eq1:ddmc}
    \frac{1}{c}\frac{\partial\phi_{0,i}}{\partial t} 
    + \frac{1}{\Delta x_i}\left(\beta_{i+1/2}\phi_{0,i+1/2} - \beta_{i-1/2}\phi_{0,i-1/2}\right)
    + \eta_i\phi_{0,i}
    + \frac{1}{\gamma_i}\sigma_{a,0,i}\phi_{0,i}
    \\
    - \frac{1}{\gamma_i^2\Delta x_i}\left(
    \frac{1}{3\gamma_{i+1/2}\sigma_{t,0,i+1/2}}\left.\frac{\partial\phi_0}{\partial x}\right|_{i+1/2}
    - \frac{1}{3\gamma_{i-1/2}\sigma_{t,0,i-1/2}}\left.\frac{\partial\phi_0}{\partial x}\right|_{i-1/2}
    \right) 
    = \frac{Q_{0,i}}{\gamma_i} \;\;,
\end{multline}
where we have dropped the superscript $(0)$, understanding that our leading-order solution will be
the full solution for the diffusion equation.
Consequently, we need to find $\phi_{0,i\pm 1/2}$ in terms of $\phi_{0,i}$ and $\phi_{0,i\pm1}$
from auxiliary equations.

A simple way to treat the advection portion of the Lagrangian time-derivative is to upwind
$\phi_{0,i\pm1/2}$ based on the sign of $\beta_{i\pm1/2}$,
\begin{equation}
    \label{eq2:ddmc}
    \phi_{0,i+1/2} = \Theta(\beta_{i+1/2})\phi_{0,i} + (1 - \Theta(\beta_{i+1/2}))\phi_{0,i+1} \;\;,
\end{equation}
where $\Theta(\cdot)$ is the standard unit step (or Heaviside) function.
Equation~\eqref{eq2:ddmc} acts either like a sink ($\beta_{i+1/2} > 0$) or a source ($\beta_{i+1/2} < 0$),
but not both.
If we permit discontinuous velocity at cell edges, introducing superscript $\pm$ to denote evaluation
immediately below (-) or above (+) the edge along $x$, we may generalize Eq.~\eqref{eq2:ddmc},
\begin{subequations}
    \label{eq3:ddmc}
    \begin{gather}
        \beta_{i+1/2}\phi_{0,i+1/2} = \Theta(\beta_{i+1/2}^{-})\beta_{i+1/2}^{-}\phi_{0,i}
        + \Theta(-\beta_{i+1/2}^{+})\beta_{i+1/2}^{+}\phi_{0,i+1} \;\;, \\
        \beta_{i-1/2}\phi_{0,i-1/2} = \Theta(-\beta_{i-1/2}^{+})\beta_{i-1/2}^{+}\phi_{0,i}
        + \Theta(\beta_{i-1/2}^{-})\beta_{i-1/2}^{-}\phi_{0,i-1} \;\;,
    \end{gather}
\end{subequations}
which permits simultaneous sources and sinks at the cell edge $i+1/2$, and reduces to
Eq.~\eqref{eq2:ddmc} when the velocity is continuous ($\beta_{i+1/2}^{-}=\beta_{i+1/2}^{+}=\beta_{i+1/2}$).
Incorporating Eq.~\eqref{eq3:ddmc} into the advection term in Eq.~\eqref{eq1:ddmc},
\begin{multline}
    \frac{1}{\Delta x_i}\left(\beta_{i+1/2}\phi_{0,i+1/2} - \beta_{i-1/2}\phi_{0,i-1/2}\right)
    = \\
    \frac{1}{\Delta x_i}\left(\Theta(\beta_{i+1/2}^{-})\beta_{i+1/2}^{-}\phi_{0,i}
    - \Theta(-\beta_{i+1/2}^{+})|\beta_{i+1/2}^{+}|\phi_{0,i+1}
    + \Theta(-\beta_{i-1/2}^{+})|\beta_{i-1/2}^{+}|\phi_{0,i}
    - \Theta(\beta_{i-1/2}^{-})\beta_{i-1/2}^{-}\phi_{0,i-1}
    \right)
    \\
    = (\sigma_{A,i\rightarrow i+1} + \sigma_{A,i\rightarrow i-1})\phi_{0,i}
    - \frac{\Delta x_{i+1}}{\Delta x_i}\sigma_{A,i+1\rightarrow i}\phi_{0,i+1}
    - \frac{\Delta x_{i-1}}{\Delta x_i}\sigma_{A,i-1\rightarrow i}\phi_{0,i-1} \;\;,
\end{multline}
where
\begin{subequations}
    \label{eq4:ddmc}
    \begin{gather}
        \sigma_{A,i\rightarrow i-1} = \frac{1}{\Delta x_i}\Theta(-\beta_{i-1/2}^{+})|\beta_{i-1/2}^{+}| \;\;, \\
        \sigma_{A,i\rightarrow i+1} = \frac{1}{\Delta x_i}\Theta(\beta_{i+1/2}^{-})\beta_{i+1/2}^{-} \;\;,
    \end{gather}
\end{subequations}
can be viewed as advection ``leakage opacities''~\cite{densmore2007} (see~\cite{abdikamalov2012} for an
operator-split continuous treatment of advection).

The diffusion operator terms can be evaluated in the usual way in the domain interior~\cite{densmore2007},
\begin{equation}
    \frac{1}{3\gamma_{i+1/2}\sigma_{t,0,i+1/2}}\left.\frac{\partial\phi_0}{\partial x}\right|_{i+1/2}
    = \frac{2(\phi_{0,i+1} - \phi_{0,i})}
    {3(\gamma_{i+1/2}^{-}\sigma_{t,0,i+1/2}^{-}\Delta x_i + \gamma_{i+1/2}^{+}\sigma_{t,0,i+1/2}^{+}\Delta x_{i+1})}
    \;\;,
\end{equation}
where $\pm$-superscripts again indicate evaluation just to the lower or upper side of the cell edge $i+1/2$.
The formulation permits several options for discretizing the velocity field, for instance: if $\gamma$
is continuous then $\gamma_{i+1/2}^{-} = \gamma_{i+1/2}^{+} = \gamma_{i+1/2}$; if $\gamma$
is piecewise-constant, then one may set $\gamma_{i+1/2}^{-} = \gamma_{i}$ and $\gamma_{i+1/2}^{+} = \gamma_{i+1}$.
Consequently, the diffusion leakage opacities, corresponding to the standard non-relativistic 
form~\cite{densmore2007}, are
\begin{subequations}
    \label{eq5:ddmc}
    \begin{gather}
        \sigma_{D,i\rightarrow i-1} = \frac{2}{3\gamma_i^2\Delta x_i}
        \left(\frac{1}{\gamma_{i-1/2}^{+}\sigma_{t,0,i-1/2}^{+}\Delta x_i
        + \gamma_{i-1/2}^{-}\sigma_{t,0,i-1/2}^{-}\Delta x_{i-1}}\right) \;\;, \\
        \sigma_{D,i\rightarrow i+1} = \frac{2}{3\gamma_i^2\Delta x_i}
        \left(\frac{1}{\gamma_{i+1/2}^{-}\sigma_{t,0,i+1/2}^{-}\Delta x_i
        + \gamma_{i+1/2}^{+}\sigma_{t,0,i+1/2}^{+}\Delta x_{i+1}}\right)
        \;\;.
    \end{gather}
\end{subequations}

Incorporating Eqs.~\eqref{eq4:ddmc} and~\eqref{eq5:ddmc} into Eq.~\eqref{eq1:ddmc},
\begin{multline}
    \label{eq6:ddmc}
    \frac{1}{c}\frac{\partial\phi_{0,i}}{\partial t}
    + \left(\sigma_{A,i\rightarrow i+1} + \sigma_{A,i\rightarrow i-1} + 
    \sigma_{D,i\rightarrow i+1} + \sigma_{D,i\rightarrow i-1} +
    \eta_i + \frac{1}{\gamma_i}\sigma_{a,0,i}\right)\phi_{0,i}
    \\
    = \frac{Q_{0,i}}{\gamma_i} + \frac{\Delta x_{i+1}}{\Delta x_i}\left(\sigma_{A,i+1\rightarrow i}
    + \frac{\gamma_{i+1}^2}{\gamma_i^2}\sigma_{D,i+1\rightarrow i}\right)\phi_{0,i+1}
    + \frac{\Delta x_{i-1}}{\Delta x_i}\left(\sigma_{A,i-1\rightarrow i}
    + \frac{\gamma_{i-1}^2}{\gamma_i^2}\sigma_{D,i-1\rightarrow i}\right)\phi_{0,i-1}
    \;\;.
\end{multline}
Equation~\eqref{eq6:ddmc} is very similar to a standard grey diffusion stencil, but with
the addition of advection leakage, the $\eta$ coefficient, and $\gamma$ factors multiplying
diffusion leakage opacities.

Finally, to treat the boundary, we assume the asymptotic boundary layer condition used
by~\cite{densmore2007} holds in the comoving frame,
\begin{equation}
    \label{eq7:ddmc}
    2\int_0^1W(\mu_0)\psi_0^-(x_{i-1/2},\mu_0)d\mu_0 = \phi_0(x_{i-1/2})
    - \frac{\lambda}{\gamma\sigma_{t,0,i-1/2}}
    \left.\frac{\partial\phi_0}{\partial x}\right|_{i-1/2} \;\;,
\end{equation}
where $W(\cdot)$ is the angular weighting function for intensity incident just below
$x_{i-1/2}$, $\psi_0^-$, and $\lambda$ is the extrapolation distance~\cite{habetler1975}.
Equation~\eqref{eq7:ddmc} is consistent with applying scaling Eqs.~\eqref{eq28:asym} to
the transport equation, keeping terms up to O($\varepsilon$), transforming intensity and
opacity to the comoving frame, and then applying the boundary layer analysis of
\cite{habetler1975} (the factor of $\gamma$ from the transformation of $\sigma_t$ to
$\sigma_{t,0}$ can be included in the transformation of the spatial coordinate to the
stretched optical depth parameter).
Equation~\eqref{eq7:ddmc} can also be obtained from the the static boundary
condition specified in comoving space-time, assuming the comoving time derivative is 0.
Following the discretization procedure of~\cite{densmore2007}, the leakage opacity and
boundary transmission probability are
\begin{subequations}
    \begin{gather}
        \tilde{\sigma}_{D,i\rightarrow i-1} = 
        \frac{2}{3\gamma_i^2\Delta x_i(\gamma_{i-1/2}^+\sigma_{t,0,i-1/2}^+\Delta x_i + 2\lambda)}
        \;\;, \\
        P_{i-1/2}(\mu_0) 
        = \frac{4}{3\gamma_i^2(\gamma_{i-1/2}^+\sigma_{t,0,i-1/2}^+\Delta x_i + 2\lambda)}
        \left(1 + \frac{3}{2}\mu_0\right) \;\;,
    \end{gather}
\end{subequations}
which provide the boundary condition stencil for the full equation,
\begin{multline}
    \frac{1}{c}\frac{\partial\phi_{0,i}}{\partial t}
    + \left(\sigma_{A,i\rightarrow i+1} + \sigma_{A,i\rightarrow i-1} + 
    \sigma_{D,i\rightarrow i+1} + \tilde{\sigma}_{D,i\rightarrow i-1} +
    \eta_i + \frac{1}{\gamma_i}\sigma_{a,0,i}\right)\phi_{0,i}
    \\
    = \frac{Q_{0,i}}{\gamma_i} 
    + \frac{\Delta x_{i+1}}{\Delta x_i}\left(\sigma_{A,i+1\rightarrow i}
    + \frac{\gamma_{i+1}^2}{\gamma_i^2}\sigma_{D,i+1\rightarrow i}\right)\phi_{0,i+1} \\
    + \frac{1}{\Delta x_i}\left(\beta_{i-1/2}^{+}\phi_{0,i-1/2}^-
    + \int_0^1\mu_0P_{i-1/2}(\mu_0)\psi_0^-(x_{i-1/2},\mu_0)d\mu_0\right)
    \;\;.
\end{multline}
In a hybrid transport-diffusion scheme, a transporting MC particle incident at $i-1/2$ on
the DDMC region can have its energy and direction transformed into the comoving frame,
then admission into the DDMC region can be determined by $P_{i-1/2}(\mu_0)$.
We note that the $\gamma$-factors in the denominator act only to lower the probability
relative to the non-relativistic form, and hence do not complicate the restriction
$P_{i-1/2}(\mu_0) \leq 1$.

\subsection{Time stencil and stability}

Considering the spatial stencil for the domain interior, and further assuming velocity
and opacity are constant in a time step, a simple time discretization is
\begin{multline}
    \label{eq1:time}
    \frac{\phi_{0,i,n+1}-\phi_{0,i,n}}{c\Delta t_n}
    + \left(\sigma_{A,i\rightarrow i+1} + \sigma_{A,i\rightarrow i-1} + 
    \sigma_{D,i\rightarrow i+1} + \sigma_{D,i\rightarrow i-1} +
    \eta_i + \frac{1}{\gamma_i}\sigma_{a,0,i}\right)\bar{\phi}_{0,i}
    \\
    = \frac{\bar{Q}_{0,i}}{\gamma_i} + \frac{\Delta x_{i+1}}{\Delta x_i}\left(\sigma_{A,i+1\rightarrow i}
    + \frac{\gamma_{i+1}^2}{\gamma_i^2}\sigma_{D,i+1\rightarrow i}\right)\bar{\phi}_{0,i+1}
    + \frac{\Delta x_{i-1}}{\Delta x_i}\left(\sigma_{A,i-1\rightarrow i}
    + \frac{\gamma_{i-1}^2}{\gamma_i^2}\sigma_{D,i-1\rightarrow i}\right)\bar{\phi}_{0,i-1}
    \;\;,
\end{multline}
where
\begin{equation}
    \label{eqA:time}
    \bar{\phi}_{0,i} = (1-\alpha)\phi_{0,i,n} + \alpha\phi_{0,i,n+1} \;\;,
\end{equation}
and $\alpha\in[0,1]$ is a time-centering parameter (consequently, $\alpha=0,1/2,1$ correspond
to explicit Euler, Crank-Nicolson, and implicit Euler schemes, respectively).
We may rewrite Eq.~\eqref{eq1:time} as
\begin{equation}
    \label{eq2:time}
    \phi_{0,i,n+1} + \mathbf{A}_{i,i-1}\bar{\phi}_{0,i-1}
    + \mathbf{A}_{i,i}\bar{\phi}_{0,i} + \mathbf{A}_{i,i+1}\bar{\phi}_{0,i+1}
    = \phi_{0,i,n} + c\Delta t_n\frac{\bar{Q}_{0,i}}{\gamma_i} \;\;,
\end{equation}
where
\begin{subequations}
    \begin{gather}
        \mathbf{A}_{i,i} = c\Delta t_n\left(\sigma_{A,i\rightarrow i+1} + \sigma_{A,i\rightarrow i-1} + 
        \sigma_{D,i\rightarrow i+1} + \sigma_{D,i\rightarrow i-1} +
        \eta_i + \frac{1}{\gamma_i}\sigma_{a,0,i}\right) \;\;, \\
        \mathbf{A}_{i,i\pm1} = -\frac{\Delta x_{i\pm1}}{\Delta x_i}
        c\Delta t_n\left(\sigma_{A,i\pm1\rightarrow i}
        + \frac{\gamma_{i\pm1}^2}{\gamma_i^2}\sigma_{D,i\pm1\rightarrow i}\right) \;\;,
    \end{gather}
\end{subequations}
are entries of a tri-diagonal matrix, which due to $\eta_i$ may not be positive-definite.
For uniform spatial cells, the $\mathbf{A}$-matrix is symmetric if 
$\beta_{i+1/2}^-=\beta_{i+1/2}^+$.
Substituting in the right side of Eq.~\eqref{eqA:time} for $\bar{\phi}_0$,
\begin{multline}
    \label{eq3:time}
    \alpha\mathbf{A}_{i,i-1}\phi_{0,i-1,n+1}
    + \left(1 + \alpha\mathbf{A}_{i,i}\right)\phi_{0,i,n+1}
    + \alpha\mathbf{A}_{i,i+1}\phi_{0,i+1,n}
    \\
    = c\Delta t_n\frac{\bar{Q}_{0,i}}{\gamma_i}
    - (1-\alpha)\mathbf{A}_{i,i-1}\phi_{0,i-1,n}
    + \left(1 - (1-\alpha)\mathbf{A}_{i,i}\right)\phi_{0,i,n}
    - (1-\alpha)\mathbf{A}_{i,i+1}\phi_{0,i+1,n}
    \;\;.
\end{multline}

We may perform a Fourier error analysis for Eq.~\eqref{eq3:time}.
For a spatial wavenumber $K$, assuming $\phi_{0,i,n}^{(e)}$ is an exact
solution to the stencil, substituting
\begin{equation}
    \phi_{0,i,n} = \phi_{0,i,n}^{(e)} + \delta\phi_{K,n}e^{iKx_i} \;\;,
\end{equation}
into Eq.~\eqref{eq3:time} gives
\begin{equation}
    \label{eq4:time}
    \frac{\delta\phi_{K,n+1}}{\delta\phi_{K,n}}
    = \frac{-(1-\alpha)\mathbf{A}_{i,i-1}e^{-iK\Delta x_{i-1/2}} + 1 - (1-\alpha)\mathbf{A}_{i,i}
    - (1-\alpha)\mathbf{A}_{i,i+1}e^{iK\Delta x_{i+1/2}}}{\alpha\mathbf{A}_{i,i-1}e^{-iK\Delta x_{i-1/2}}
    + 1 + \alpha\mathbf{A}_{i,i}
    + \alpha\mathbf{A}_{i,i+1}e^{iK\Delta x_{i+1/2}}}
    \;\;.
\end{equation}
We note we have abused notation by using $i$ as an imaginary number when multiplying $K$, and as
a cell index subscript.
The stability condition is that the modulus of Eq.~\eqref{eq4:time} be less than or equal to 1.
In Section~\ref{sec:numres}, we only use a uniform spatial grid; setting
$\theta = K\Delta x = K\Delta x_{i\pm1/2}$ in Eq.~\eqref{eq4:time},
\begin{equation}
    \label{eq5:time}
    \frac{\delta\phi_{K,n+1}}{\delta\phi_{K,n}}
    = \frac{\mathcal{R}(\phi,\alpha-1)}{\mathcal{R}(\phi,\alpha)}
    \;\;,
\end{equation}
where
\begin{equation}
    \mathcal{R}(\theta,z) = 1 + z\mathbf{A}_{i,i} +
    z\mathbf{A}_{i,i-1}e^{-i\theta} + z\mathbf{A}_{i,i+1}e^{i\theta} \;\;.
\end{equation}

\begin{figure}[t]
    \centering
    \includegraphics[width=0.49\linewidth]{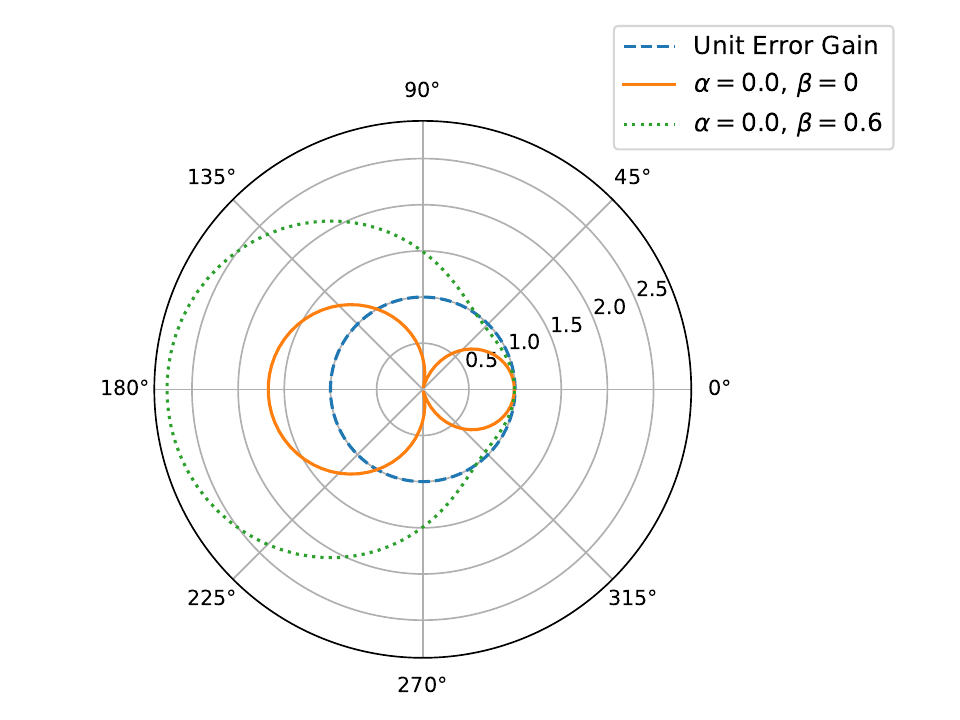}
    \includegraphics[width=0.49\linewidth]{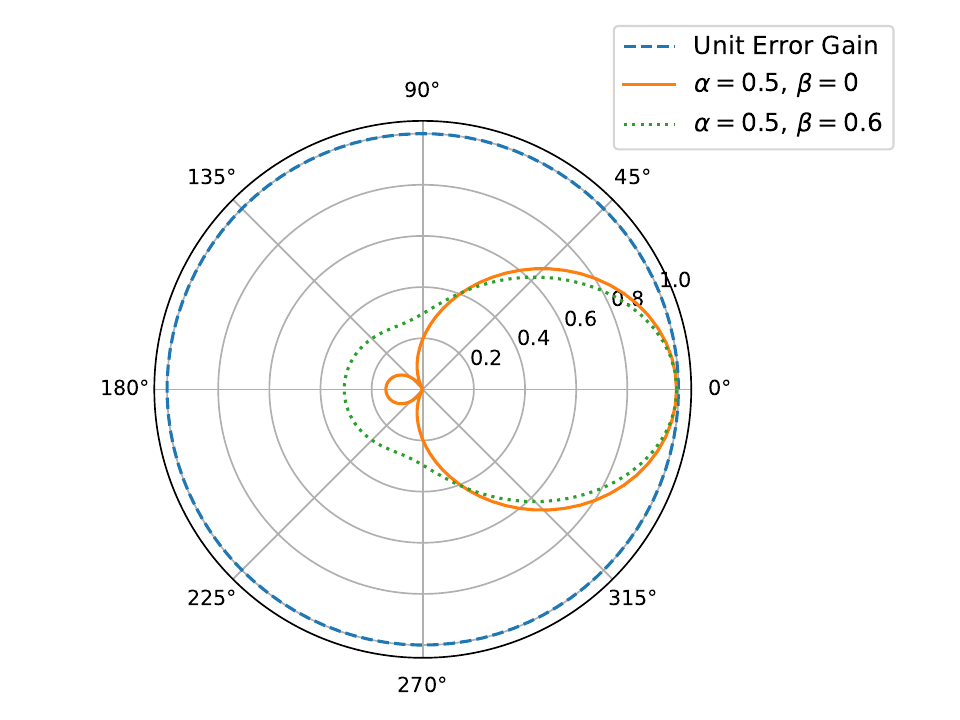}
    \caption{Polar graphs of stability measure $\mathcal{G}(\theta,\alpha)$ for $\beta=0.6$
    versus angle $\theta=K\Delta x$ (green dotted).
    Time centering is $\alpha=0$ (explicit Euler; left panel) or $\alpha=0.5$ (Crank-Nicolson;
    right panel).
    For comparison, the unit circle (blue dashed) and $\mathcal{G}(\theta,\alpha)$ with
    $\beta=0$ (orange solid) are also plotted.}
    \label{fg1:time}
\end{figure}

Figure~\ref{fg1:time} has example stability curves for $\beta=0.6$ and $\alpha=0$ (explicit Euler)
or $\alpha=0.5$ (Crank-Nicolson), with $\Delta x, c\Delta t, \sigma_t,\sigma_a=1/128,1/64,128,1/2$,
respectively, and
$\mathcal{G}(\theta,\alpha) = |\mathcal{R}(\phi,\alpha-1)/\mathcal{R}(\phi,\alpha)|$.
Comparing to the static diffusion equation ($\beta=0$), we see the effect of time centering
on stability depends on velocity: for $\alpha=0$ there is a significantly larger range of spatial
wavenumbers that correspond to unstable error modes.
The most significant contributor to the enhancement in instability for $\alpha=0$ are the advective
terms resulting from the Lagrangian derivative, consistent with advection imposing a CFL-type condition
on time step size.

As the right panel of Fig.~\ref{fg1:time} suggests, increasing the time-centering parameter makes
the time-evolution of the error more stable; for the parameters considered the stability profile
of the drift-diffusion equation is comparable to that of the static diffusion equation.
Furthermore, $\alpha=0.5$ should furnish 2nd-order accuracy in time.
In Section~\ref{sec:numres}, we consider time-independent code implementations for comparing
accuracy of the diffusion equation to lab-frame Monte Carlo transport; thus we avoid the issue
of the time-stability considered in the present section.

\section{Numerical results}
\label{sec:numres}

In order to gauge accuracy of the drift-diffusion equation relative to transport, we consider several
simple steady velocity profiles; we also provide the corresponding semi-relativistic drift-diffusion
solutions to Eq.~\eqref{eq24:asym}.
In particular, in Section~\ref{sec:constbeta} we examine the solutions for several uniform constant
$\beta$ profiles, removing the effect of adiabatic expansion and Doppler shift.
In Section~\ref{sec:linebeta}, We test a few values of the velocity gradient, for a velocity linearly
proportional to $x$ (sometimes referred to as ``homologous flow'' in supernova literature).
Given the local time derivative of velocity is 0 in these problems, in principle the spatial velocity
gradient must be small to warrant application of Eq.~\eqref{eq26:asym}.
Finally, in Section~\ref{sec:jumpbeta} we examine a jump discontinuity that breaks velocity continuity
at a spatial cell edge.

Our reference solution is a special-relativistic Monte Carlo transport implementation, where particle
motion, or streaming, is performed in the lab frame while particle interactions are performed in the
comoving frame.
We employ a linear sub-cell profile for velocity, so the particle distance to collision is
\begin{equation}
    d_p = -\frac{\ln(\xi)}{(\sigma_{t,p}-\sigma_{a,p})}=
    -\frac{\ln(\xi)}{\gamma_p(1-\mu\beta_p)(\sigma_{t,0}-\sigma_{a,0})} \;\;,
\end{equation}
where subscript $p$ denotes evaluation of the quantity at the location of the particle, $x_p$,
\begin{subequations}
    \begin{gather}
        \beta_p = (\beta_{i+1/2}^--\beta_{i-1/2}^+)\frac{(x_p - x_{i-1/2})}{\Delta x_i} + \beta_{i-1/2}^+
        \;\;,\\
        \gamma_p = \frac{1}{\sqrt{1-\beta_p^2}} \;\;.
    \end{gather}
\end{subequations}
We note that this distance formula does not take into account the change in velocity over the particle
path traversing distance $d$.
In order to obtain the comoving scalar intensity from the Monte Carlo, we tally a path-length estimator
\begin{equation}
    \phi_{0,i,{\rm MC}} = \frac{c}{\Delta x_i}\sum_p
    \sum_{d_p\in i}\gamma_p^2(1-\beta_p\mu_p)^2E_p\frac{(1-e^{-\sigma_{a,p}d_p})}{\sigma_{a,p}} \;\;,
\end{equation}
where $d_p\in i$ implies the sum is over the set of tracks of particle $p$ inside cell $i$.
We note that expanding the quadratic and distributing the sum furnishes path length estimators
for lab-frame energy density, flux and pressure in the context of the standard energy-momentum
tensor transformation for comoving energy density (see, for instance,~\cite{castor2004},
Chapter 6).

For the following calculations, we fix the spatial domain $x\in[0,1]$ with 128 spatial cells.
We also fix the comoving opacities as $\sigma_{t,0}=128$ and $\sigma_{a,0}=1/2$.
All problems have vacuum boundary conditions on both sides of the spatial domain.
Units are omitted as the linear transport and diffusion equations are spatially scale-free,
so $\sigma_{t,0}=128$ and $\partial\beta/\partial x = 0.03$ are 128 mean-free paths and
a change in $\beta$ by 0.03 over some unit of length (e.g.~the radius of a star). 
For the MC transport, we employ 512 particles per cell for the uniform volume source and
65,536 particles for the point source.
All comparisons in this section use lab-frame time-independent
code implementations of both the discrete diffusion and Monte Carlo.

\subsection{Constant $\beta$}
\label{sec:constbeta}

\begin{figure}[t]
    \centering
    \includegraphics[width=0.6\linewidth]{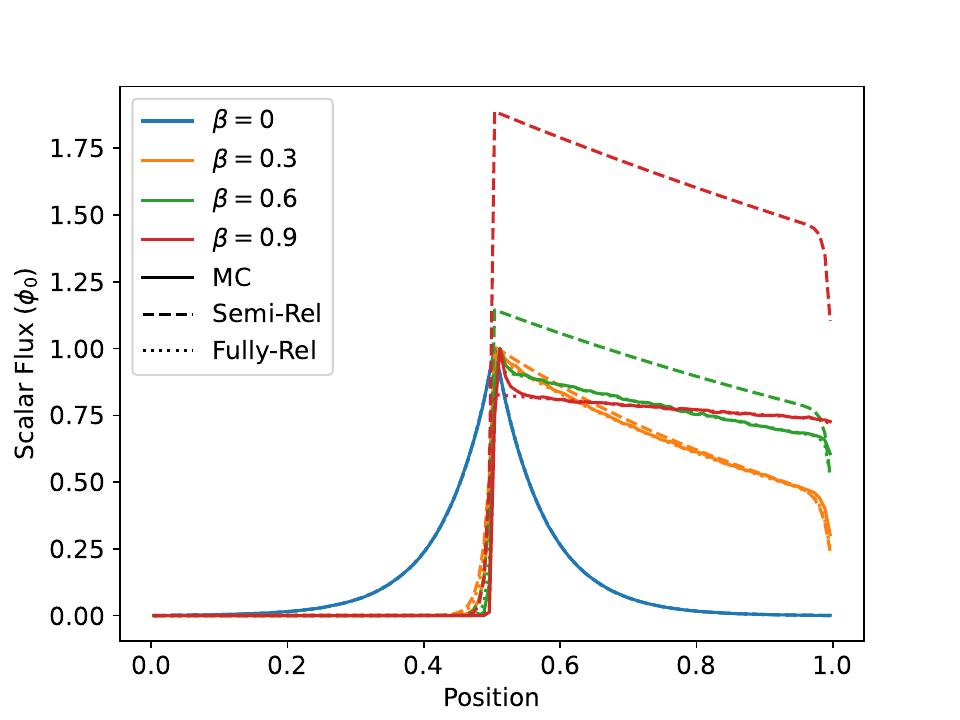}
    \caption{Scaled scalar flux versus spatial coordinate for a point source at
    $x=0.5$ and several values of constant $\beta$.
    The maximum value of MC transport (solid) is set to 1, and semi-relativistic (dashed)
    and fully relativistic diffusion (dotted) are scaled accordingly.}
    \label{fg1:constbeta}
\end{figure}

Here we compare the drift-diffusion equation with lab frame MC transport at several
constant velocities.
This problem has a discretized point source $Q_{0,i} = 1/\Delta x$ at $x=0.5$.
This removes the effect of the adiabatic and Doppler shift term in the drift diffusion
equation, isolating the effect of the $\gamma$-factors.
Figure~\ref{fg1:constbeta} has scaled comoving scalar flux versus position, where
each MC (solid lines) result has its peak normalized to 1, and the corresponding diffusion
results are scaled by the same resulting factor, per $\beta$ value.
The result for $\beta=0$ merely confirms the static solutions are consistent.
At $\beta=0.3$, we see the effect of beaming in the MC and the effect of advection in
diffusion cause the solutions to skew to the right.
Moreover, at $\beta=0.3$, all three results agree closely.
However, at $\beta=0.6$ and 0.9, we see that the semi-relativistic result suffers in
comparison to the MC result, consistent with O($\beta^2$) effects from the $\gamma$-factors
becoming important.
The fully relativistic drift-diffusion result agrees closely at each $\beta$, but is
missing a spike near the source present in the MC result.

\subsection{Homologous flow, $\beta = \beta_c + (x-1/2)\Delta\beta$}
\label{sec:linebeta}

\begin{figure}[t]
    \centering
    \includegraphics[width=0.6\linewidth]{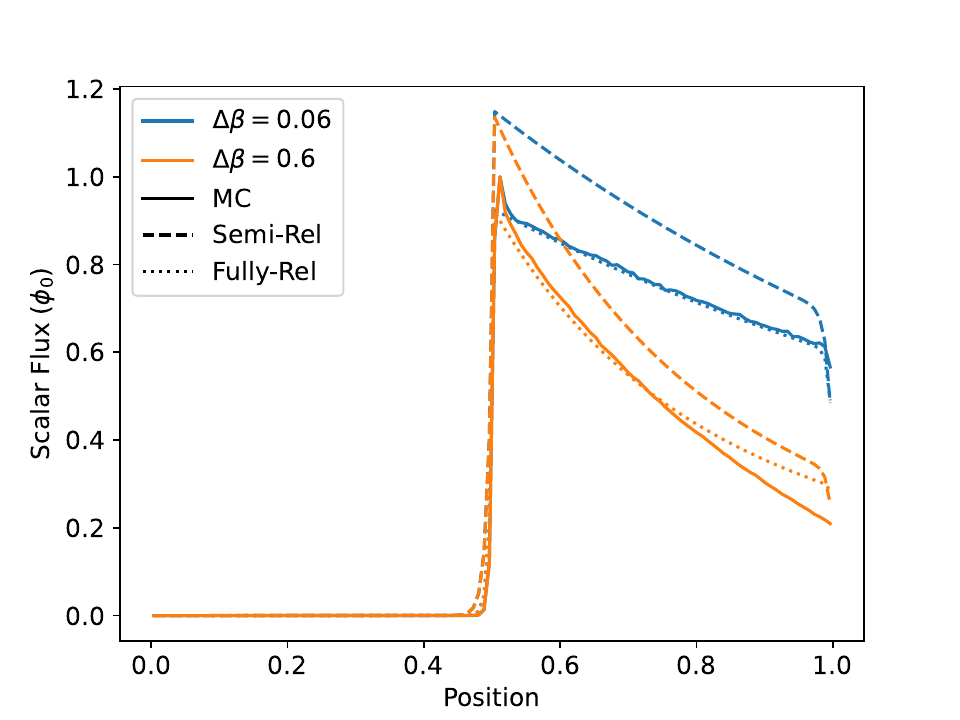}
    \caption{Scaled scalar flux versus spatial coordinate for a point source at
    $x=0.5$ and several values of $\Delta\beta=0.06,0.6$ in $\beta = 0.6 + (x-1/2)\Delta\beta$.
    The maximum value of MC transport (solid) is set to 1, and semi-relativistic (dashed)
    and fully relativistic diffusion (dotted) are scaled accordingly.}
    \label{fg1:linebeta}
\end{figure}

We next test a linear profile of the form $\beta = 0.6 + (x-1/2)\Delta\beta$,
where $\Delta\beta = 0.06$ or 0.6.
Otherwise, we preserve the settings from Section~\ref{sec:constbeta}.
Figure~\ref{fg1:linebeta} has comoving scaler flux versus position for the two values
for MC (solid lines), semi-relativistic diffusion (dashed lines), and fully-relativistic
diffusion (dotted lines).
The fully relativistic solution again gives better agreement to the MC transport result.
However, for $\Delta\beta=0.6$ we see some discrepancy towards the right boundary, where
the fully-relativistic diffusion solution becomes closer to that of the semi-relativistic
diffusion.
It is possible the degree of anisotropy near the higher bound in the MC solution is not
captured in the asymptotic boundary condition we have implemented.

\subsection{Velocity jump, $\beta = \beta_l+\Theta(x-1/2)\Delta\beta$}
\label{sec:jumpbeta}

\begin{figure}[t]
    \centering
    \includegraphics[width=0.6\linewidth]{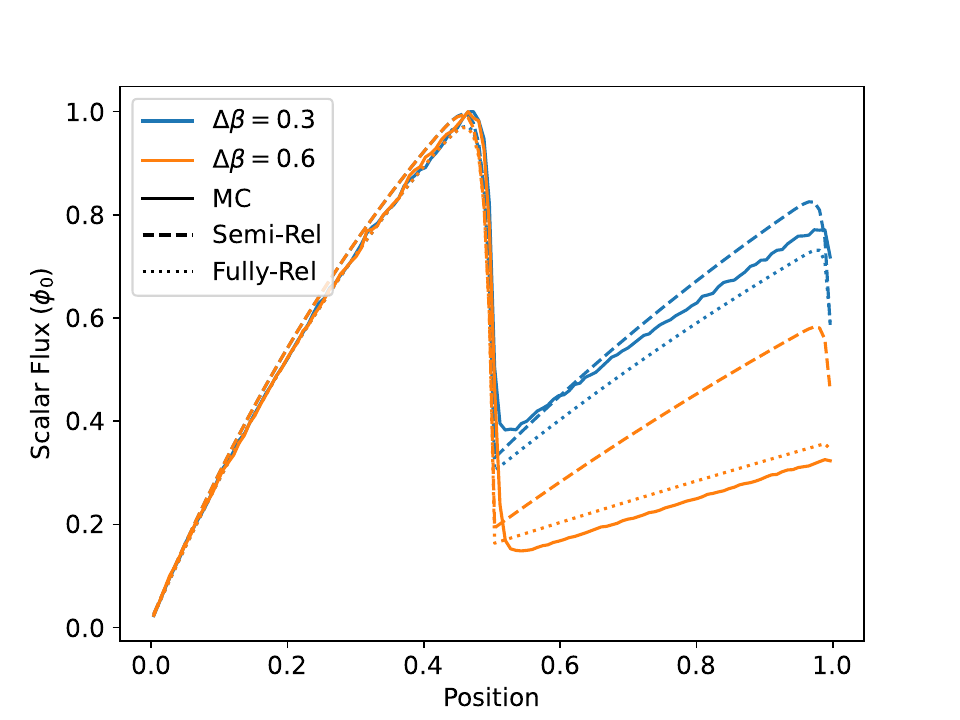}
    \caption{Scaled scalar flux versus spatial coordinate for a uniform comoving
    source and several values of $\Delta\beta=0.06,0.6$ in $\beta = 0.3 + \Theta(x-1/2)\Delta\beta$.
    The maximum value of MC transport (solid) is set to 1, and semi-relativistic (dashed)
    and fully relativistic diffusion (dotted) are scaled accordingly.}
    \label{fg1:jumpbeta}
\end{figure}

Finally, we consider a problem with a very sharp, localized velocity gradient,
where our velocity discretization approximates $\beta = 0.3+\Theta(x-1/2)\Delta\beta$,
where $\Theta(\cdot)$ is again the unit step function.
Departing from the previous problems we consider a uniform comoving volume source,
$Q_0=1$.
The jump in velocity at $x=1/2$ incurs a significant change in the comoving scalar
intensity, as seen in Fig.~\ref{fg1:jumpbeta} for all three solutions.
In Fig.~\ref{fg1:jumpbeta} we have MC transport (solid lines), semi-relativistic (dashed lines)
and fully-relativistic (dotted lines) diffusion.
We see that all three solutions are comparable for the $\Delta\beta=0.3$ jump
(to $\beta=0.6$ at $x\geq 1/2$).
For the larger jump $\Delta\beta=0.6$, we see that the fully relativistic solution is
in better agreement in the higher-velocity region, but all three solutions agree fairly
well in the lower-velocity region.

\section{Conclusions}

We have derived a fully-relativistic grey diffusion equation for continuous-direction
1D planar linear radiative transfer using a lab-frame-based asymptotic analysis.
This analysis makes use of a family of parameter scaling that is intended to be in
proximity to the standard static-material background scaling used in linear transport
theory.
The resulting fully-relativistic diffusion equation has a diffusion operator that scales
as $1/\gamma^3$, consistent with the two-direction Poisson-Kac-based derivation
of~\cite{giona2017}.

We have scrutinized conditions for which the asymptotic analysis gives a closed form
of the diffusion equation, which are not merely the standard parameter scaling relationships
in the comoving frame; these give several pathologies.
One pathology is an occurance of 0 being equal to a non-zero term in the O($\varepsilon$)
equation, which can be avoided by rescaling the comoving time derivative plus an anisotropic
coefficient of the angular intensity.
After this adjustment, another pathology is in the closed form of the equation: without
asymptotically scaling the Lagrangian derivative of velocity, we see a parity violation
of the solution when inverting the spatial coordinate and negating the velocity (the solution
does not behave forward and backward in $x$ in the same way).
Thus we must asymptotically scale the Lagrangian derivative of the velocity (or $\beta$),
which indicates the resulting fully-relativistic diffusion equation can only be reliably
accurate when fluid parcels are nearly not accelerating (this condition nearly exactly
holds for many supernova ejecta, which follows $v=x/t$).
The fully-relativistic 1D diffusion equation is straightforwardly amenable to a DDMC treatment,
and hence hybridization with MC transport for acceleration.

We have also presented several simple numerical tests of the fully-relativistic diffusion
equation, comparing to the $\gamma$-less semi-relativistic diffusion equation commonly used
in the literature, and to a lab-frame Monte Carlo transport solution.
For the problems tested, we see that the $\gamma$-dressing of the fully-relativistic diffusion
equation furnishes an improvement in agreement to the lab-frame transport at $\beta\gtrsim0.6$,
relative to the semi-relativistic diffusion result.
All solutions have been presented for steady-state in the lab frame, but the extension to
time-dependence is amenable to standard time-discretization; however, for explicit or 
semi-implicit schemes care must be taken to select a sufficiently small time step size, due to
the CFL-type condition imposed by the advection term in the Lagrangian time derivative.

The main objectives for future work on this topic are as follows:
\begin{enumerate}
    \item thermal radiative transfer,
    \item hybridizing with Monte Carlo transport,
    \item extending to 3D and non-planar geometries,
    \item and incorporating frequency dependence.
\end{enumerate}
The standard static-material scaling rules for thermal radiative transfer are somewhat
different than linear transport: the absorption opacity is scaled to be asymptotically
large and one considers a matter equation with an asymptotically small heat capacity
\cite{larsen1983}.
The time derivative is scaled as in the linear transport parameter scaling.
This approach is evidently distinct from the opacity-only scaling method of
\cite{thomas1930,anderson1972} considered in the general-relativistic framework.
General relativistic codes often use a time increment that does not correspond to a simple
lab-frame picture, however, and the time variable is on more equal footing with the
spatial variables (see Appendix Section~\ref{sec:thomas}).

For 3D, we must consider the finding of~\cite{giona2017} that shows an anisotropic effective
diffusion coefficient.
While this result is still for discrete directions, we anticipate a similar effect for
continuous direction in 3D.
One possible path forward on a 3D derivation is to consider the analysis on a piecewise
constant velocity field, where one can rotate the spatial coordinate system to have an
axis aligned with the velocity in each cell; then it may be possible to follow much of
the analysis given here, but with an additional step of inverting the spatial rotation
(and projecting it onto a stencil that is compatible with the cells).

\section{Acknowledgments}
\label{sec:ack}

We thank Nick Gentile for useful discussions.
This work has been assigned document release number LA-UR-25-31726.
This work was supported by the U.S. Department of Energy through the Los Alamos National
Laboratory.
Los Alamos National Laboratory is operated by Triad National Security, LLC, for the National
Nuclear Security Administration of U.S. Department of Energy (Contract No. 89233218CNA000001).

\begin{appendices}

\section{$\Lambda$-integral recursion}
\label{sec:app1}

Integrals of $\lambda_{n,k}$ over $\mu$ take the form,
\begin{equation}
    \label{eq1:app1}
    \Lambda_{n,k} = \int_{-1}^1\left(\frac{1}{\gamma(1-\beta\mu)}\right)^n\mu^kd\mu 
    = \frac{1}{\gamma^n}\int_{-1}^1\left(\frac{1}{1-\beta\mu}\right)^n\mu^kd\mu \;\;,
\end{equation}
where $n$ and $k$ are assumed to be non-negative integers.
Substituting $y = 1 - \beta\mu$,
\begin{multline}
    \label{eq2:app1}
    \Lambda_{n,k} = \frac{1}{\gamma^n}\int_{1+\beta}^{1-\beta}
    \frac{1}{y^n}\left(\frac{1-y}{\beta}\right)^k\left(-\frac{1}{\beta}\right)dy 
    = \frac{1}{\gamma^n\beta^{k+1}}\int_{1-\beta}^{1+\beta}\frac{1}{y^n}(1-y)^kdy
    \\
    = \frac{1}{\gamma^n\beta^{k+1}}\int_{1-\beta}^{1+\beta}
    \frac{1}{y^n}\sum_{j=0}^k{k \choose j}(-y)^jdy
    = \frac{1}{\gamma^n\beta^{k+1}}\sum_{j=0}^k{k \choose j}(-1)^j
    \int_{1-\beta}^{1+\beta}y^{j-n}dy
    \;\;.
\end{multline}
The rightmost integral evaluates to
\begin{equation}
    \label{eq3:app1}
    \int_{1-\beta}^{1+\beta}y^{j-n}dy
    = \begin{cases}
        \displaystyle
        \frac{1}{j-n+1}\left((1+\beta)^{j-n+1} - (1-\beta)^{j-n+1}\right) \;\;,\;\; j\not=n-1 \;\;,
        \\\\
        \displaystyle
        \ln\left(\frac{1+\beta}{1-\beta}\right) \;\;,\;\; j=n-1 \;\;.
    \end{cases}
\end{equation}
If $j < n - 1$,
\begin{multline}
    \label{eq4:app1}
    \int_{1-\beta}^{1+\beta}y^{j-n}dy
    = \frac{1}{n-1-j}\left(\frac{1}{(1-\beta)^{n-1-j}} - \frac{1}{(1+\beta)^{n-1-j}}\right)
    \\
    = \frac{\gamma^{2(n-1-j)}}{n-1-j}\left((1+\beta)^{n-1-j} - (1-\beta)^{n-1-j}\right)
    = \frac{2\gamma^{2(n-1-j)}}{n-1-j}\sum_{l=0}^{2l+1\leq n-1-j}{n-1-j\choose 2l+1}\beta^{2l+1}
    \;\;.
\end{multline}

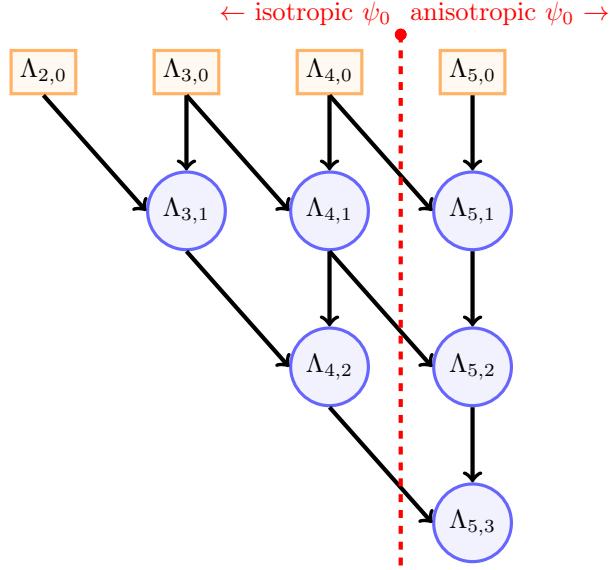
\begin{figure}
    \centering
    \begin{tikzpicture}[
        roundnode/.style={circle, draw=blue!60, fill=blue!5, very thick, minimum size=7mm},
        squarednode/.style={rectangle, draw=orange!60, fill=orange!5, very thick, minimum size=5mm},
        ]

        \node[squarednode](lam20){$\Lambda_{2,0}$};
        \node[squarednode](lam30)[right=of lam20]{$\Lambda_{3,0}$};
        \node[squarednode](lam40)[right=of lam30]{$\Lambda_{4,0}$};
        \node[squarednode](lam50)[right=of lam40]{$\Lambda_{5,0}$};
        \node[roundnode](lam31)[below=of lam30]{$\Lambda_{3,1}$};
        \node[roundnode](lam41)[below=of lam40]{$\Lambda_{4,1}$};
        \node[roundnode](lam51)[below=of lam50]{$\Lambda_{5,1}$};
        \node[roundnode](lam42)[below=of lam41]{$\Lambda_{4,2}$};
        \node[roundnode](lam52)[below=of lam51]{$\Lambda_{5,2}$};
        \node[roundnode](lam53)[below=of lam52]{$\Lambda_{5,3}$};

        \draw[ultra thick, ->] (lam20.south) -- (lam31.west);
        \draw[ultra thick, ->] (lam30.south) -- (lam31.north);
        \draw[ultra thick, ->] (lam30.south) -- (lam41.west);
        \draw[ultra thick, ->] (lam40.south) -- (lam41.north);
        \draw[ultra thick, ->] (lam40.south) -- (lam51.west);
        \draw[ultra thick, ->] (lam50.south) -- (lam51.north);
        \draw[ultra thick, ->] (lam31.south) -- (lam42.west);
        \draw[ultra thick, ->] (lam41.south) -- (lam42.north);
        \draw[ultra thick, ->] (lam41.south) -- (lam52.west);
        \draw[ultra thick, ->] (lam51.south) -- (lam52.north);
        \draw[ultra thick, ->] (lam42.south) -- (lam53.west);
        \draw[ultra thick, ->] (lam52.south) -- (lam53.north);

        \filldraw[red] (4.75,0.5) circle (2pt) node[anchor=south east]{$\leftarrow$ isotropic $\psi_0$};
        \filldraw[red] (4.75,0.5) circle (2pt) node[anchor=south west]{anisotropic $\psi_0$ $\rightarrow$};
        \draw[ultra thick, dashed, red, -] (4.75,0.5) -- (4.75,-6.6);
    \end{tikzpicture}
    \caption{Example diagram of triangular recursion given by Eq.~\eqref{eq7:app1}.
    The first layer of orange rectangles is evaluated with Eqs.~\eqref{eq2:app1} and
    \eqref{eq4:app1}.
    This particular diagram generates the factors needed for evaluating lab-frame
    $\mu$-weighted integrals from the corresponding comoving $\mu_0$-weighted integrals,
    up to linear anisotropy in the comoving frame.
    The red dashed line separates the factors needed for isotropic comoving intensity
    (left) from those needed for linear anisotropic comoving intensity (right).}
    \label{fg1:app1}
\end{figure}

A corollary of Eq.~\eqref{eq2:app1} is $\Lambda_{n,k}$ is determined by $\Lambda_{n-j,0}$ for
$j\in\{0,\ldots,k\}$; given
\begin{equation}
    \label{eq5:app1}
    \int_{1-\beta}^{1+\beta}y^{j-n}dy = 
    \sum_{j'=0}^0{0\choose j'}(-1)^{j'}\int_{1-\beta}^{1+\beta}y^{j'-(n-j)}dy
    = \gamma^{n-j}\beta\Lambda_{n-j,0} \;\;,
\end{equation}
then
\begin{equation}
    \label{eq6:app1}
    \Lambda_{n,k} = \frac{1}{\gamma^n\beta^k}\sum_{j=0}^k{k \choose j}(-1)^j
    \gamma^{n-j}\Lambda_{n-j,0}
    \;\;.
\end{equation}
For $n$ and $k$ greater than 1,
\begin{equation}
    \label{eq7:app1}
    \Lambda_{n,k} = \frac{1}{\gamma^n\beta^k}
    \left(-\gamma^{n-1}\beta^{k-1}\Lambda_{n-1,k-1} + \gamma^n\beta^{k-1}\Lambda_{n,k-1}\right)
    = \frac{1}{\beta}\Lambda_{n,k-1} - \frac{1}{\gamma\beta}\Lambda_{n-1,k-1}
    \;\;,
\end{equation}
which follows from
\begin{equation}
    \label{eq8:app1}
    \int\frac{y^k}{(1-y)^n}dy = \int\frac{y^k-y^{k-1}}{(1-y)^n}dy + \int\frac{y^{k-1}}{(1-y)^n}dy
    = -\int\frac{y^{k-1}}{(1-y)^{n-1}}dy + \int\frac{y^{k-1}}{(1-y)^n}dy
\end{equation}
(it can also be derived from Eq.~\eqref{eq6:app1} using recursion of binomial coefficients).
Figure~\ref{fg1:app1} shows an example diagram of the recursion given by Eq.~\eqref{eq7:app1},
relevant to calculations of lab-frame quantities from different $\mu_0$-expansions of comoving
intensity, $\psi_0$, evaluated in the following sections.

Evaluating the $k=0$ integrals with Eq.~\eqref{eq4:app1}, and using the recursion pattern
in Fig.~\ref{fg1:app1} gives Table~\ref{tb1:app1}.
Notably, odd values of $k$ correspond to terms with lowest power in $\beta$ greater than 0,
so that they vanish when $\beta=0$, as expected by inspection of Eq.~\eqref{eq1:app1}.
\begin{table}[ht]
    \centering
    \begingroup
    \renewcommand{\arraystretch}{1.5}
    \begin{tabular}{c|ccc:c}
        k \textbackslash~n & 2 & 3 & 4 & 5 \\
        \hline
        0 & $2$ & $2\gamma$ & $\frac{2\gamma^2}{3}(3+\beta^2)$ & $2\gamma^3(1+\beta^2)$ \\
        1 & - & $2\gamma\beta$ & $\frac{8\gamma^2}{3}\beta$ & $\frac{2\gamma^3}{3}\beta(5+\beta^2)$ \\
        2 & - & - & $\frac{2\gamma^2}{3}(1+3\beta^2)$ & $\frac{2\gamma^3}{3}(1+5\beta^2)$ \\
        3 & - & - & - & $2\gamma^3\beta(1+\beta^2)$ \\
    \end{tabular}
    \endgroup
    \caption{Evaluation of recursion diagram in Fig.~\ref{fg1:app1}, using Eq.~\eqref{eq4:app1}
        for $k=0$ and Eq.~\eqref{eq7:app1} for $k>0$.}
    \label{tb1:app1}
\end{table}

\section{Lab-frame P$_1$ expansion}
\label{sec:app2}

For convenience, we rewrite the lab-frame transport equation here:
\begin{equation}
    \label{eq1:lfgme}
    \frac{1}{c}\frac{\partial\psi}{\partial t}
    + \mu\frac{\partial \psi}{\partial x} + \sigma_{t,0}\gamma(1-\beta\mu)\psi = 
    \frac{1}{2}\left(\frac{\sigma_{s,0}}{\gamma(1-\beta\mu)^3}\right)
    \int_{-1}^1(1-\beta\mu')^2\psi(\mu')d\mu' + \frac{q}{2} \;\;.
\end{equation}
Defining
\begin{subequations}
    \label{eq2:lfgme}
    \begin{gather}
        E = \frac{1}{c}\int_{-1}^1\psi d\mu \;\;, \\
        F = \int_{-1}^1\mu\psi d\mu \;\;, \\
        P = \frac{1}{c}\int_{-1}^1\mu^2\psi d\mu \;\;, \\
        Q_E = \frac{1}{2}\int_{-1}^1q d\mu \;\;, \\
        Q_M = \frac{1}{2}\int_{-1}^1\mu q d\mu \;\;,
    \end{gather}
\end{subequations}
the 0th and 1st moment angular integrals of Eq.~\eqref{eq1:lfgme} are
\begin{subequations}
    \label{eq3:lfgme}
    \begin{gather}
         \frac{\partial E}{\partial t}
        + \frac{\partial F}{\partial x} + \sigma_{t,0}\gamma(cE-\beta F) = 
        \sigma_{s,0}\gamma^3(cE - 2\beta F + c\beta^2P) + Q_E \;\;, \\
        \frac{1}{c}\frac{\partial F}{\partial t}
        + c\frac{\partial P}{\partial x} + \sigma_{t,0}\gamma(F-c\beta P) = 
        \sigma_{s,0}\gamma^3\beta(cE - 2\beta F + c\beta^2P) + Q_M \;\;.
    \end{gather}
\end{subequations}

\subsection{Lab frame for isotropic comoving intensity}
\label{sec:cmviso}

If the comoving frequency-integrated intensity is isotropic,
\begin{equation}
    \label{eq1:cmviso}
    \psi_0 = \int_0^{\infty}\psi_{0,\nu_0}d\nu_0 = \frac{cE_0}{2} \;\;,
\end{equation}
then the lab-frame intensity satisfies
\begin{multline}
    \label{eq2:cmviso}
    \psi = \int_0^{\infty}\psi_{\nu}d\nu 
    = \int_0^{\infty}\left(\frac{\nu}{\nu_0}\right)^3\psi_{0,\nu_0}d\nu
    = \int_0^{\infty}\left(\frac{\nu}{\nu_0}\right)^3\psi_{0,\nu_0}\left(\frac{\nu}{\nu_0}\right)d\nu_0
    \\
    = \left(\frac{1}{\gamma(1-\beta\mu)}\right)^4\int_0^{\infty}\psi_{0,\nu_0}d\nu_0
    = \left(\frac{1}{\gamma(1-\beta\mu)}\right)^4\psi_0
    = \left(\frac{1}{\gamma(1-\beta\mu)}\right)^4\frac{cE_0}{2} \;\;.
\end{multline}

The lab-frame energy density, flux, and pressure are the 0th, 1st, and 2nd moments in lab-frame angle $\mu$,
\begin{subequations}
    \begin{gather}
        \label{eq3:cmviso}
        E = \frac{1}{c}\int_{-1}^1\psi d\mu = \frac{E_0}{2}\Lambda_{4,0} \;\;,\\
        F = \int_{-1}^1\mu\psi d\mu = \frac{cE_0}{2}\Lambda_{4,1} \;\;,\\
        P = \frac{1}{c}\int_{-1}^1\mu^2\psi d\mu = \frac{E_0}{2}\Lambda_{4,2} \;\;.
    \end{gather}
\end{subequations}
Using the values from Table~\ref{tb1:app1},
\begin{subequations}
    \label{eq5:cmviso}
    \begin{gather}
        E = \frac{E_0}{2}\Lambda_{4,0} = \gamma^2\left(\frac{3+\beta^2}{3}\right)E_0 \;\;,\\
        F = \frac{cE_0}{2}\Lambda_{4,1} = \gamma^2\frac{4}{3}c\beta E_0 \;\;,\\
        P = \frac{E_0}{2}\Lambda_{4,2} = \gamma^2\left(\frac{1+3\beta^2}{3}\right)E_0 \;\;.
    \end{gather}
\end{subequations}
We can now incorporate these identities into Eq.~\eqref{eq3:lfgme}a, after simplification
\begin{equation}
    \label{eq6:cmviso}
    \frac{\partial}{\partial t}\left(\gamma^2\left(\frac{3+\beta^2}{3}\right)E_0\right)
    + \frac{\partial}{\partial x}\left(\gamma^2\frac{4}{3}c\beta E_0\right)
    + \sigma_{t,0}\gamma cE_0 = 
    \sigma_{s,0}\gamma cE_0 + Q_E \;\;,
\end{equation}
which can be re-written terms of $E$, $F$, or $P$ as the only dependent variable.
This equation describes the effects of advection, expansion, absorption (subtracting the
scattering term from the total attenuation), and Doppler shift (in grey-integrated form),
given some source $Q_E$.
Writing in terms of $E$ and cancelling the scattering term,
\begin{equation}
    \label{eq7:cmviso}
    \frac{\partial E}{\partial t}
    + \frac{\partial}{\partial x}\left(\frac{4c\beta}{3+\beta^2}E\right)
    + \frac{1}{\gamma}\sigma_{a,0}\left(\frac{3}{3 + \beta^2}\right)cE = Q_E \;\;.
\end{equation}
Taking the limit as $\beta\rightarrow 1$ of this equation,
\begin{equation}
    \label{eq8:cmviso}
    \frac{\partial E}{\partial t} + c\frac{\partial E}{\partial x}
    = Q_E \;\;.
\end{equation}
Thus at near-light speed, an isotropic comoving solution corresponds to
a simple light-speed advection lab-frame solution, along $x$.
It is also straightforward to show the lab-frame signal speed implied by Eq.~\eqref{eq7:cmviso}
is causally bounded:
\begin{equation*}
    \frac{4|\beta|}{3+\beta^2} \leq 1 \iff 0 \leq \beta^2 - 4|\beta| + 3 = (3-|\beta|)(1-|\beta|) \;\;.
\end{equation*}

\subsection{Lab frame for linearly anisotropic comoving intensity}
\label{sec:cmvani}

Here we generalize Eq.~\eqref{eq1:cmviso},
\begin{equation}
    \label{eq1:cmvani}
    \psi_0 = \int_0^{\infty}\psi_{0,\nu_0}d\nu_0 = \frac{cE_0}{2} + \frac{3}{2}\mu_0F_0 \;\;.
\end{equation}
Thus $F_0$ is the comoving flux.
In 1D planar geometry the Lorentz transform of direction reduces to
\begin{equation}
    \label{eq2:cmvani}
    \mu_0 = \frac{\mu - \beta}{1 - \beta\mu} \;\;,
\end{equation}
so the lab-frame intensity is
\begin{multline}
    \label{eq3:cmvani}
    \psi = \left(\frac{1}{\gamma(1-\beta\mu)}\right)^4\psi_0
    = \left(\frac{1}{\gamma(1-\beta\mu)}\right)^4\left(\frac{cE_0}{2}
    + \frac{3}{2}\left(\frac{\mu - \beta}{1 - \beta\mu}\right)F_0\right)
    \\
    = \tilde{\psi} 
    + \frac{3\gamma}{2}\left(\frac{1}{\gamma(1-\beta\mu)}\right)^5
    \left(\mu - \beta\right)F_0 \;\;,
\end{multline}
where $\tilde{\phi}$ has been introduced to account for the contribution of the comoving
isotropic intensity term to the lab-frame intensity.
The lab-frame energy density, flux, and pressure
\begin{subequations}
    \label{eq4:cmvani}
    \begin{gather}
        E = \tilde{E} + \frac{3\gamma}{2c}F_0\int_{-1}^1
        \left(\frac{1}{\gamma(1-\beta\mu)}\right)^5\left(\mu - \beta\right)d\mu 
        = \tilde{E} + \frac{3\gamma}{2c}F_0\left(\Lambda_{5,1}-\beta\Lambda_{5,0}\right) \;\;,\\
        F = \tilde{F} + \frac{3\gamma}{2}F_0\left(\Lambda_{5,2}-\beta\Lambda_{5,1}\right) \;\;,\\
        P = \tilde{P} + \frac{3\gamma}{2c}F_0\left(\Lambda_{5,3}-\beta\Lambda_{5,2}\right) \;\;.
    \end{gather}
\end{subequations}
where $\tilde{E}$, $\tilde{F}$ and $\tilde{P}$ are the moments of $\tilde{\psi}$, and
are given by Eq.~\eqref{eq5:cmviso}.
Using the rightmost column of Table~\ref{tb1:app1}, the $\Lambda$-coefficients in
Eqs.~\eqref{eq4:cmvani} become
\begin{subequations}
    \label{eq6:cmvani}
    \begin{gather}
        \Lambda_{5,1} - \beta\Lambda_{5,0} = 
        2\gamma^3\left(\frac{1}{3}\beta(5+\beta^2) - \beta(1+\beta^2)\right) 
        = \frac{4\gamma^3}{3}\beta\left(1 - \beta^2\right) = \frac{4\gamma}{3}\beta
        \;\;,\\
        \Lambda_{5,2} - \beta\Lambda_{5,1} 
        = 2\gamma^3\left(\frac{1}{3}\left(1+5\beta^2\right)
        - \frac{1}{3}\beta^2(5+\beta^2)\right) 
        = \frac{2\gamma^3}{3}\left(1+5\beta^2 - \beta^2(5+\beta^2)\right)
        = \frac{2\gamma^3}{3}\left(1 - \beta^4\right)
        \nonumber\\
        = \frac{2\gamma}{3}(1+\beta^2)
        \;\;,\\
        \Lambda_{5,3} - \beta\Lambda_{5,2} = 2\gamma^3\left(
        \beta(1+\beta^2) - \beta \frac{1}{3}\left(1+5\beta^2\right)\right)
        = \frac{2\gamma^3}{3}\beta\left(3+3\beta^2 - 1-5\beta^2\right)
        = \frac{4\gamma^3}{3}\beta\left(1-\beta^2\right)
        \nonumber\\
        = \frac{4\gamma}{3}\beta \;\;.
    \end{gather}
\end{subequations}
Evaluating Eqs.~\eqref{eq4:cmvani} with Eqs.~\eqref{eq6:cmvani}
\begin{subequations}
    \label{eq7:cmvani}
    \begin{gather}
        E = \tilde{E} + \frac{3\gamma}{2c}F_0\left(\Lambda_{5,1}-\beta\Lambda_{5,0}\right)
        = \tilde{E} + 2\gamma^2\frac{1}{c}F_0\beta
        = \gamma^2\left(\left(\frac{3+\beta^2}{3}\right)E_0 + 2\frac{\beta}{c}F_0\right) \;\;,\\
        F = \tilde{F} + \frac{3\gamma}{2}F_0\left(\Lambda_{5,2}-\beta\Lambda_{5,1}\right)
        = \tilde{F} + \gamma^2F_0(1+\beta^2) 
        = \gamma^2\left(\frac{4}{3}c\beta E_0 + F_0(1+\beta^2)\right) \;\;,\\
        P = \tilde{P} + \frac{3\gamma}{2c}F_0\left(\Lambda_{5,3}-\beta\Lambda_{5,2}\right)
        = \tilde{P} + 2\gamma^2\frac{1}{c}F_0\beta
        = \gamma^2\left(\left(\frac{1+3\beta^2}{3}\right)E_0 + 2\frac{\beta}{c}F_0\right) \;\;.
    \end{gather}
\end{subequations}
We see that for $\beta=0$, $F=F_0$, as expected.
The linear aniostropic contribution to comoving intensity adds the same contribution to
$\tilde{E}$ and $\tilde{P}$ to obtain $E$ and $P$, respectively.
We may write the equations for $E$ and $F$ as
\begin{equation}
    \label{eq8:cmvani}
    \left(
    \begin{array}{c}
        E \\
        F 
    \end{array}
    \right)
    =
    \gamma^2\left(
    \begin{array}{cc}
        \frac{3+\beta^2}{3} & 2\frac{\beta}{c} \\
        \frac{4}{3}c\beta & 1+\beta^2
    \end{array}
    \right)
    \left(
    \begin{array}{c}
        E_0 \\
        F_0 
    \end{array}
    \right) \;\;.
\end{equation}
Inverting the 2x2 matrix,
\begin{equation}
    \label{eq9:cmvani}
    \left(
    \begin{array}{c}
        E_0 \\
        F_0 
    \end{array}
    \right)
    =
    \frac{3}{\gamma^2(3-4\beta^2+\beta^4)}
    \left(
    \begin{array}{cc}
        1+\beta^2 & -2\frac{\beta}{c} \\
        -\frac{4}{3}c\beta & \frac{3+\beta^2}{3}
    \end{array}
    \right)
    \left(
    \begin{array}{c}
        E \\
        F 
    \end{array}
    \right)
    =
    \frac{3}{(3-\beta^2)}
    \left(
    \begin{array}{cc}
        1+\beta^2 & -2\frac{\beta}{c} \\
        -\frac{4}{3}c\beta & \frac{3+\beta^2}{3}
    \end{array}
    \right)
    \left(
    \begin{array}{c}
        E \\
        F 
    \end{array}
    \right)
\end{equation}
It can be seen that $\beta=0$ reduces the 2x2 matrix to the identity
matrix, in Eqs.~\eqref{eq8:cmvani} and~\eqref{eq9:cmvani}.
Using Eq.~\eqref{eq9:cmvani}, $P$ can be expressed in terms of $E$ and $F$,
\begin{multline}
    \label{eq10:cmvani}
    P = \gamma^2\frac{3}{(3-\beta^2)}\left(
    \left(\frac{1+3\beta^2}{3}\right)
    \left((1+\beta^2)E-2\frac{\beta}{c}F\right)
    + 2\frac{\beta}{c}
    \left(\frac{3+\beta^2}{3}F-\frac{4}{3}c\beta E\right)\right)
    \\
    = \frac{\gamma^2}{(3-\beta^2)}\left(
    ((1+3\beta^2)(1+\beta^2)-8\beta^2)E
    + ((3+\beta^2)-(1+3\beta^2))2\frac{\beta}{c}F
    \right)
    \\
    = \frac{\gamma^2}{(3-\beta^2)}\left(
    (1-\beta^2)(1-3\beta^2)E + 4(1-\beta^2)\frac{\beta}{c}F\right)
    = \frac{1}{(3-\beta^2)}\left(
    (1-3\beta^2)E + 4\frac{\beta}{c}F\right)
\end{multline}
Equation~\eqref{eq10:cmvani} has the following limits in $\beta$,
\begin{subequations}
    \label{eq11:cmvani}   
    \begin{gather}
        \lim_{\beta\rightarrow 0}P = \frac{E}{3} \;\;, \\
        \lim_{\beta\rightarrow 1}P = \frac{2}{c}F-E \;\;,
    \end{gather}
\end{subequations}
where the $\beta\rightarrow 0$ limit is the expected static isotropic pressure.
The $\beta\rightarrow 1$ limit is consistent with (but not derived from)
$\psi = cE\delta(\mu-1)$, where $\delta(\cdot)$ is the Dirac delta distribution.

Augmenting Eqs.~\eqref{eq3:lfgme} with Eq.~\eqref{eq10:cmvani}, the closed
system of equations is
\begin{subequations}
    \label{eq12:cmvani}
    \begin{gather}
         \frac{\partial E}{\partial t}
        + \frac{\partial F}{\partial x} + \sigma_{t,0}\gamma(cE-\beta F) = 
        \sigma_{s,0}\gamma^3(cE - 2\beta F + c\beta^2P) + Q_E \;\;, \\
        \frac{1}{c}\frac{\partial F}{\partial t}
        + c\frac{\partial P}{\partial x} + \sigma_{t,0}\gamma(F-c\beta P) = 
        \sigma_{s,0}\gamma^3\beta(cE - 2\beta F + c\beta^2P) + Q_M \;\;, \\
        P = \frac{1}{(3-\beta^2)}\left((1-3\beta^2)E + 4\frac{\beta}{c}F\right)
        \;\;.
    \end{gather}
\end{subequations}

\subsection{O($\beta$) and neglecting derivatives of $F$ and $\beta F$}
\label{sec:obeta}

\subsubsection{O($\beta$)}

To O($\beta$), Eqs.~\eqref{eq12:cmvani} become
\begin{subequations}
    \label{eq1:obeta}
    \begin{gather}
         \frac{\partial E}{\partial t}
        + \frac{\partial F}{\partial x} + \sigma_{t,0}(cE-\beta F) = 
        \sigma_{s,0}(cE - 2\beta F) + Q_E \;\;, \\
        \frac{1}{c}\frac{\partial F}{\partial t}
        + c\frac{\partial P}{\partial x} + \sigma_{t,0}(F-c\beta P) = 
        c\sigma_{s,0}\beta E + Q_M \;\;, \\
        P = \frac{1}{3}\left(E + 4\frac{\beta}{c}F\right)
        \;\;.
    \end{gather}
\end{subequations}
Incorporating Eq.~\eqref{eq1:obeta}c into Eq.~\eqref{eq1:obeta}b and
neglecting O($\beta^2$),
\begin{equation}
    \label{eq2:obeta}
    \frac{1}{c}\frac{\partial F}{\partial t}
    + \frac{4}{3}\frac{\partial}{\partial x}(\beta F)
    + \frac{c}{3}\frac{\partial E}{\partial x}
    + \sigma_{t,0}F-\frac{1}{3}c\beta\sigma_{t,0}E = 
    c\beta\sigma_{s,0}E + Q_M \;\;.
\end{equation}
This can be re-written in terms of the comoving time derivative
(at O($\beta$) now equivalent to the Lagrangian time derivative),
\begin{equation}
    \label{eq3:obeta}
    \frac{1}{c}\frac{\partial F}{\partial t_0}
    + F\frac{\partial\beta}{\partial x}
    + \frac{1}{3}\frac{\partial}{\partial x}(\beta F)
    + \frac{c}{3}\frac{\partial E}{\partial x}
    + \sigma_{t,0}F-\frac{1}{3}c\beta\sigma_{t,0}E = 
    c\beta\sigma_{s,0}E + Q_M \;\;.
\end{equation}

\subsubsection{Neglecting the comoving time derivative of $F$}

If we neglect the comoving time derivative of $F$ and simplify,
\begin{equation}
    \label{eq4:obeta}
    \left(\frac{4}{3}\frac{\partial\beta}{\partial x} + \sigma_{t,0}\right)F
    + \frac{\beta}{3}\frac{\partial F}{\partial x}
    = -\frac{c}{3}\frac{\partial E}{\partial x}
    + c\beta\left(\frac{1}{3}\sigma_{t,0} + \sigma_{s,0}\right)E + Q_M \;\;.
\end{equation}
Equation~\eqref{eq4:obeta} is a 1st order ordinary differential equation for $F$.
Assuming $\beta\not=0$, the solution is
\begin{multline}
    \label{eq5:obeta}
    F(x) = \left(\frac{\beta(x)}{\beta(0)}\right)^4e^{-3\int_0^x\sigma_{t,0}/\beta(x') dx'}F(0) \\
    + \int_0^x\left(\frac{\beta(x')}{\beta(0)}\right)^4e^{-3\int_{x'}^x\sigma_{t,0}/\beta(x'') dx''}
    \left(-\frac{c}{\beta(x')}\frac{\partial E}{\partial x'} + c(\sigma_{t,0}+3\sigma_{s,0})E(x')
    + \frac{3}{\beta(x')}Q_M(x')\right)dx'
\end{multline}
where we have taken $x=0$ to be the integration bound.
If we take $\beta$ to be constant, then take the limit as $\beta$ goes to 0, then in
the second term on the right side of Eq.~\eqref{eq5:obeta} we have indeterminate forms
\begin{equation}
    \label{eq6:obeta}
    \lim_{\beta\rightarrow 0}\frac{e^{-3\sigma_{t,0}(x-x')/\beta}}{\beta}
    = \frac{1}{3\sigma_{t,0}}\lim_{\beta\rightarrow 0}
    \frac{3\sigma_{t,0}e^{-3\sigma_{t,0}(x-x')/\beta}}{\beta}
    = \frac{1}{3\sigma_{t,0}}\delta^{(+)}(x-x')
    \;\;,
\end{equation}
multiplying the gradient of energy density and $Q_M$, where we have introduced a half-space
Dirac delta distribution in the final equality, given the indeterminate form satisfies
the criteria of being a nascent half-space Dirac delta distribution (compact support and
unit integral).
Thus in this limit Eq.~\eqref{eq5:obeta} reduces to
\begin{equation}
    \label{eq7:obeta}
    F = -\frac{c}{3\sigma_{t,0}}\frac{\partial E}{\partial x} + \frac{1}{\sigma_{t,0}}Q_M
    \;\;,
\end{equation}
which is the static form of Fick's Law (if the transport source term is isotropic in the lab frame,
then $Q_M=0$).

\subsubsection{Neglecting terms $\sim\partial (\beta F)/\partial x$}

If the terms consisting of a spatial gradient of $F$ or $\beta$ multiplied by $\beta$ or $F$
are neglected, Eq.~\eqref{eq4:obeta} reduces to
\begin{equation}
    \label{eq8:obeta}
    F = -\frac{c}{3\sigma_{t,0}}\frac{\partial E}{\partial x}
    + c\beta\left(\frac{1}{3} + \frac{\sigma_{s,0}}{\sigma_{t,0}}\right)E
    + \frac{1}{\sigma_{t,0}}Q_M \;\;.
\end{equation}
Incorporating Eq.~\eqref{eq8:obeta} into Eq.~\eqref{eq1:obeta} and keeping O($\beta$) terms
\begin{multline}
    \label{eq9:obeta}
    \frac{\partial E}{\partial t} + c\beta\frac{\partial E}{\partial x}
    - \frac{\partial}{\partial x}\left(\frac{c}{3\sigma_{t,0}}\frac{\partial E}{\partial x}\right)
    + c\frac{\partial}{\partial x}
    \left(\left(\frac{1}{3} + \frac{\sigma_{s,0}}{\sigma_{t,0}}\right)\beta E\right)
    - \frac{2}{3}c\left(1 + \frac{\sigma_{s,0}}{\sigma_{t,0}}\right)
    \beta\frac{\partial E}{\partial x}
    + \sigma_{a,0}cE
    \\
    = Q_E + \left(1 - \frac{2\sigma_{s,0}}{\sigma_{t,0}}\right)\beta Q_M
    - \frac{\partial}{\partial x}\left(\frac{1}{\sigma_{t,0}}Q_M\right) \;\;.
\end{multline}
The comoving scattering ratio appears in several terms in Eq.~\eqref{eq9:obeta}.
Taking the limit as $\sigma_{s,0}\rightarrow\sigma_{t,0}$ and simplifying,
\begin{equation}
    \label{eq10:obeta}
    \frac{\partial E}{\partial t} + c\beta\frac{\partial E}{\partial x}
    - \frac{\partial}{\partial x}\left(\frac{c}{3\sigma_{t,0}}\frac{\partial E}{\partial x}\right)
    + \frac{4}{3}cE\frac{\partial\beta}{\partial x}
    + \sigma_{a,0}cE = Q_E - \beta Q_M
    - \frac{\partial}{\partial x}\left(\frac{1}{\sigma_{t,0}}Q_M\right) \;\;.
\end{equation}
Equation~\eqref{eq10:obeta} is the 1D planar form of Eq.~6.51 of~\cite{castor2004}, but
with the following modifications:
\begin{itemize}
    \item the gradient of $F$ is replaced by the diffusion operator on $E$ (Fick's Law),
    \item the pressure tensor is replaced by the identity matrix multiplied by $E/3$,
    \item an additional term, $-\partial(Q_M/\sigma_{t,0})/\partial x$, appears on the right side.
\end{itemize}
However, Eq.~6.51 of~\cite{castor2004} is for comoving $E$ and $F$.
Similarly, this is the limit we obtain of~\eqref{eq23:asym} (\eqref{eq24:asym}) in 
Section~\ref{sec:asym}, but with lab-frame $E$.

\section{The Thomas scaling in non-relativistic 1D planar geometry}
\label{sec:thomas}

Here we briefly consider the Thomas scaling~\cite{thomas1930}, as presented by
\cite{anderson1972}, in a non-relativistic 1D context with constant pure absorption opacity.
To the best of our understanding, the only small parameter is the mean-free path in these
approaches.
This should constitute a simplification of Section IV of~\cite{anderson1972}.
Using similar notation to~\cite{anderson1972}, the transport equation is then
\begin{equation}
    \label{eq1:thomas}
    \frac{1}{\sigma}\frac{d\psi}{d\lambda} = \frac{S}{2}-\psi \;\;,
\end{equation}
where we have written the partial derivatives of the streaming operator in terms of the
characteristic, or affine coordinate $\lambda$.
Following~\cite{anderson1972}, we write the asymptotic expansion directly in powers of
$\sigma$.
\begin{equation}
    \label{eq2:thomas}
    \psi = \sum_{k=0}^{\infty}\frac{\tilde{\psi}^{(k)}}{\sigma^k} \;\;.
\end{equation}
We note that this is equivalent to Eq.~\eqref{eq1:asym}, by introducing a length scale
$L$, and setting $\varepsilon=1/L\sigma$ and $\psi^{(k)} = L^k\tilde{\psi}^{(k)}$.
Again following~\cite{anderson1972}, matching orders to O($1/\sigma$) gives
\begin{subequations}
    \begin{gather}
        \tilde{\psi}^{(0)} = \frac{S}{2} \;\;, \\
        \tilde{\psi}^{(1)} = -\frac{d\psi^{(0)}}{ds}
        = -\frac{1}{2}\frac{dS}{d\lambda}
    \end{gather}
\end{subequations}
(corresponding to their equation 37 in Section IV).
Applying the expansion to O($1/\sigma$) in Eq.~\eqref{eq1:thomas} and simplifying gives
\begin{equation}
    \label{eq3:thomas}
    -\frac{1}{\sigma}\frac{d^2S}{d\lambda^2} = 0 \;\;.
\end{equation}
Expanding the affine derivative using the 1D planar assumption and integrating over $\mu$,
\begin{equation}
    \label{eq3:thomas}
    \frac{1}{c\sigma}\frac{d^2S}{dt^2} + \frac{1}{3\sigma}\frac{d^2S}{dx^2} = 0 \;\;,
\end{equation}
where we have assumed $S$ is isotropic.

If instead we only expand the left side of Eq.~\eqref{eq1:thomas}, then relabel
$S$ as $\psi$ on the left side, we obtain
\begin{equation}
    \label{eq4:thomas}
    \frac{d}{d\lambda}\left(\psi-\frac{1}{\sigma}\frac{d\psi}{d\lambda}\right)
    = \sigma\left(\frac{S}{2}-\psi\right) \;\;,
\end{equation}
which upon expansion into $(x,t)$ and integration over $\mu$ gives
\begin{equation}
    \label{eq5:thomas}
    \frac{1}{c}\frac{\partial\phi}{\partial t} -
    \frac{1}{\sigma}\left(\frac{1}{c^2}\frac{\partial^2}{\partial t^2}
    +\frac{1}{3}\frac{\partial^2}{\partial x^2}\right)\phi
    = \sigma(S-\phi) \;\;,
\end{equation}
which is nearly the Telegrapher's equation (it is so with an imaginary relaxation
coefficient of the 2nd derivative in time).
Putting into the 4-vector notation of~\cite{anderson1972,shibata2011},
Eq.~\eqref{eq5:thomas} becomes
\begin{equation}
    \label{eq6:thomas}
    u^{\alpha}\nabla_\alpha\phi -
    \frac{1}{3\sigma}\nabla_{\alpha}
    \left(\left(3u^{\alpha}u^{\delta}+h^{\alpha\delta}\right)\nabla_\gamma\phi\right)
    = \sigma(S-\phi) \;\;,
\end{equation}
where $\alpha$ and $\delta$ here are 4-indexes, $u^{\alpha}=(1,0,0,0)$ is the 4-velocity,
and $h^{\alpha\delta}=u^\alpha u^\delta - \eta^{\alpha\delta}$ is the rank-4 projection
operator, as in~\cite{anderson1972,shibata2011} ($\eta^{\alpha\delta}$ here is the
Minkowski metric).

In contrast to equation 5.21 of~\cite{shibata2011}, Eq.~\eqref{eq6:thomas} has an additional
time-like component in the 2nd derivative, which can be seen as a contribution of
$\tilde{\phi}^{(1)}$ (consistent with equation 5.13 of~\cite{shibata2011}).
Excising this time-like component would furnish a non-equilibrium diffusion equation
from Eqs.~\eqref{eq5:thomas} and~\eqref{eq6:thomas}.

\end{appendices}

\bibliography{refs}
\bibliographystyle{elsarticle-num}

\end{document}